\documentclass[dvipsnames]{aa}
\pdfoutput=1

\ifx\pdfoutput\undefined
\usepackage{graphicx}
\else
\usepackage[pdfauthor={J.~Threlfall},citecolor=blue,linkcolor=blue, linktocpage=true,colorlinks=true]{hyperref}
\usepackage{epstopdf}
\fi 

\usepackage{natbib}
\bibpunct{(}{)}{;}{a}{}{,} % to follow the A&A style
\usepackage{txfonts}
\usepackage{subfig}	% subfig support
\usepackage{epsfig}	% epsfigs
\usepackage{cancel}
\usepackage{multirow}
\usepackage{fixltx2e}	% figure ordering
\usepackage{amssymb}	% extra maths symbols
\graphicspath{{figs/}}

\newsavebox{\bigleftbox}

\newcommand{\beq}{ \begin{eqnarray} }
\newcommand{\eeq}{ \end{eqnarray} }

\newcommand{\vpar}{ v_{\parallel} }
\newcommand{\vperp}{ v_{\perp} }
\newcommand{\Eperp}{ E_{\perp} }

\newcommand{\boldnabla}{\mbox{\boldmath$\nabla$}}

\def\grad{\boldnabla}

\def\vscl{{v_{scl}}}
\def\vsclsq{{{v_{scl}}^2}}
\def\lscl{{l_{scl}}}
\def\tscl{{t_{scl}}}
\def\bscl{{b_{scl}}}
\def\escl{{{e_{scl}}}}

\def\KEscl{{{{KE}_{scl}}}}
\def\omscl{{{\Omega_{scl}}}}

\newcommand{\Epar}{E_{\parallel}}
\newcommand{\upar}{u_{\parallel}}
\newcommand{\ue}{{{\bf{u}}_E}}

\date{}			% Suppresses the date after the title.....

\usepackage{color}
\definecolor{darkgreen}{rgb}{0,0.45,0}

\newcommand{\new}[1]{{{{#1}}}}	% JT comments

\DeclareRobustCommand*{\unit}[1]{\def~{\,}\ensuremath{\mathrm{\,#1}}}

\begin{document}
 \title{Particle dynamics in a non-flaring solar active region {\new{model}}}
 \titlerunning{Particle dynamics in a non-flaring solar active region model}
 \author{J.~Threlfall\inst{\ref{inst1}} \and Ph.-A. Bourdin\inst{\ref{inst2}} \and T.~Neukirch\inst{\ref{inst1}} \and C.~E.~Parnell\inst{\ref{inst1}}  }
 \institute{School of Mathematics and Statistics, University of St Andrews, St Andrews, Fife, KY16 9SS, U.K. \email{\{jwt9;tn3;cep\}@st-andrews.ac.uk}\label{inst1} \and Space Research Institute, Austrian Academy of Sciences, Schmiedlstr. 6, 8042 Graz, Austria 
 \email{Philippe.Bourdin@oeaw.ac.at}\label{inst2}}
 \abstract
 {}
 {The aim of this work is to investigate and characterise particle behaviour in an (observationally-driven) 3D MHD model of the solar atmosphere above a slowly evolving, non-flaring active region.}
 {We use a relativistic guiding-centre particle code to investigate
 \new{the behaviour of selected particle orbits, distributed throughout}
 a single snapshot of the 3D MHD simulation.} 
 {\new{Two distinct particle acceleration behaviours are recovered, which affect both electrons and protons: (i) direct acceleration along field lines and (ii) tangential drifting of guiding centres with respect to local magnetic field. 
 However, up to 40\% of all particles actually 
 experience a form of (high energy) particle trap, because of changes in the direction of the electric field and unrelated to the strength of the magnetic field; such particles are included in the first category.}
 \new{Additionally, category (i) electron and proton orbits undergo surprisingly strong acceleration to non-thermal energies ($\lesssim42$\unit{MeV}), because of the strength and extent of super-Dreicer electric fields created by the MHD simulation. Reducing the electric field strength of the MHD model does not significantly affect the efficiency of the (electric field-based) trapping mechanism, but does reduce the peak energies gained by orbits. We discuss the implications for future experiments, which aim to simulate non-flaring active region heating and reconnection.}}
 {}
 \maketitle

%%%%%%%%%%%%%%%%%%%%%%%%%%%%%%%%%%%%%%%%%%%
\section{Introduction}\label{sec:Intro}
Understanding the dynamics of charged particles in reconnecting solar processes, in particular the acceleration of particles in solar flares, is of fundamental importance for advancing our knowledge of plasma behaviour in general \citep[for example, recent reviews on particle acceleration in flares include][]{review:Vilmeretal2011,review:Fletcheretal2011,review:Zharkovaetal2011,review:Cargilletal2012}. There has been growing observational evidence that accelerated particle populations also occur in smaller scale processes such as micro-flares \citep[see e.g.][and references therein]{review:Hannahetal2011} or nano-flares \citep[][]{paper:Testaetal2014}; these processes are also often associated with magnetic reconnection. 

In the context of the solar atmosphere, it is now widely accepted that magnetic reconnection contributes to the maintenance of thermal \citep[contributing to coronal heating as discussed in][]{review:ParnelldeMoortel2012,review:Parnelletal2015} and sporadic generation of non-thermal (high-energy) particle populations. Since magnetic reconnection is intrinsically associated with field aligned electric fields \citep[e.g.][]{paper:Schindleretal1988,paper:HesseSchindler1988,paper:Schindleretal1991} and these electric fields are well-known accelerators of charged particles, the question of how the local particle dynamics on all scales are affected by these reconnection electric fields remains open and important. Additionally, magnetic reconnection can lead to rapid magnetic field changes that can also cause particle acceleration, such as collapsing magnetic traps \citep[e.g.][]{paper:GradyNeukirch2009, paper:Gradyetal2012, paper:EradatOskouietal2014,paper:EradatOskouiNeukirch2014}.

Previous studies of particle behaviour typically use a test particle approach to study a wide variety of configurations, all of which are underpinned by or directly investigate magnetic reconnection. For instance, the acceleration of particles at specific topological features has been explored in studies at/near 2/2.5D null points \citep[e.g.][]{paper:BulanovSasorov1976,paper:BruhwilerZweibel1992,paper:Kliem1994,paper:Litvinenko1996,paper:BrowningVekstein2001,paper:ZharkovaGordovskyy2004,paper:ZharkovaGordovskyy2005,paper:WoodNeukirch2005,paper:HannahFletcher2006,paper:Drakeetal2006} and, more recently, 3D null points (e.g. \citealp[][]{paper:DallaBrowning2005,paper:DallaBrowning2006,paper:DallaBrowning2008,paper:Guoetal2010,paper:Stanieretal2012}), stressed magnetic fields/current sheets \citep[e.g.][]{paper:Turkmanietal2005,paper:Onofrietal2006,paper:Gordovskyyetal2010b} and 3D magnetic separators \citep{paper:Threlfalletal2015,paper:Threlfalletal2015b}. Acceleration in more complex 3D models, such as twisted coronal loops \citep{paper:GordovskyyBrowning2011,paper:Gordovskyyetal2014}, have also been investigated. 
A significant step forward from the analysis of a localised reconnection site was taken by \citet{paper:Baumannetal2013}, who combined localised particle-in-cell (PIC) analysis in the vicinity of a reconnecting null-point with a magnetohydrodynamic (MHD) model of the surrounding active region (AR). This allowed the global impact sites of the particles to be identified.

Instead of considering an isolated reconnection process, whether embedded in a global field or not, we take a different approach and investigate the particle acceleration in an observationally driven MHD simulation of an AR. A decade ago, the first MHD simulations of downscaled ARs were produced that were based on extrapolated observed magnetograms, driven in a manner which approximated motions arising from convection cells on a range of scales \citep[e.g.][]{paper:Gudiksen2005b,paper:Gudiksen2005a}. Only recently have large-scale computations been performed that model a full AR without downscaling \citep{paper:Bourdin+al:2013_overview}.

The MHD simulation we will use was designed to model the coronal response to photospheric driving by advection of field-lines resulting from horizontal granular motions in the photosphere, where the plasma beta is greater than one \citep{paper:Bourdin+al:2013_overview}. The base of the simulation is prescribed by an observed AR magnetogram with the subsequent field-line braiding inducing currents in the corona caused by quasi-static changes propagating at the Alfv{\'e}n speed along the field into the corona. In effect, this is one implementation of the field-line braiding mechanism of \citet{paper:Parker1972} and results in coronal loops with temperatures of around 1.5\unit{MK}. The location, apex height, and plasma flows along these loops all broadly match observations \citep{paper:Bourdin+al:2013_overview}. The feasibility of this field-line braiding mechanism to heat the corona sufficiently by Ohmic dissipation of induced currents is demonstrated in \citet{paper:Bourdin+al:2015_energy-input}. This suggests that reconnection occurs regularly throughout the AR, and, hence, there must be widespread parallel electric fields. 
 
In this paper, we investigate what, if any, particle acceleration occurs in this MHD simulation of an observed non-flaring AR \citep{paper:Bourdin+al:2013_overview}. In this kind of model there are a wide range of locations (magnetic configurations) at which reconnection occurs, as well as a range of different reconnection regimes in operation. Thus, the MHD model we investigate here incorporates many of the particle acceleration scenarios that have been considered in isolation, as mentioned above. Thus, in this paper we consider the electromagnetic fields that have been generated in a non-flaring active-region simulation based on observations and determine the effects they have on charged particle dynamics.  
\begin{figure*}[t]
 \centering
 \resizebox{.98\textwidth}{!}{\includegraphics{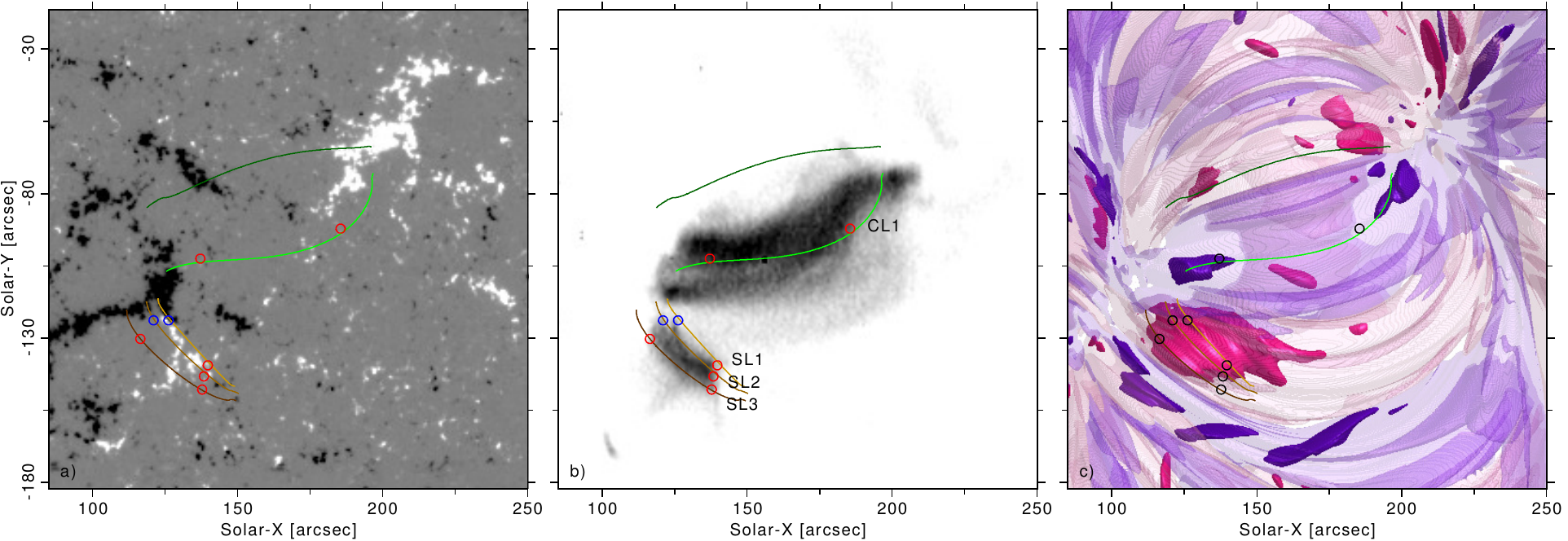}}
 \caption{Panels, from L-R, showing a) a Hinode/SOT magnetogram of an AR core seen on the 14th Nov 2007 saturated at $\pm300$\unit{G}, b) a Hinode/EIS image of the \ion{Fe}{xv} emission (at $1.5\unit{MK}$) of the corona above this region and c) the co-spatial and co-temporal simulated parallel electric field strength and orientation within this region (where pink/purple represents positive/negative $E_{||}$, and the opacity represents the relative strength of $E_{||}$ - fully opaque regions identify \new{$E_{||}=0.25E_{||}^{\rm{peak}}$ while transparent regions show $E_{||}=0.01E_{||}^{\rm{peak}}$, for a peak coronal electric field strength $E_{||}^{\rm{peak}}\sim3.5\unit{Vm^{-1}}$ in the simulation domain)}. The right hand panel is generated by simulations described in \citet{paper:Bourdin+al:2013_overview}, which also describes field lines which cross the intensity maxima of several EUV-emissive coronal loops, which have been overlaid in each panel; "CL" indicates core loops, while "SL" indicates shorter loops. The circles indicate EUV-emissive loop footpoints in the model and the corresponding plasma motion (and observed Doppler shifts) at these locations; red represents draining (i.e. motion along the line-of-sight away from an observer) while blue represents upflows (motion towards an observer).}
 \label{fig:Eparconfig}
\end{figure*}

The primary objective of the present work is to survey the particle dynamics which result from MHD simulations of a non-flaring AR, and to determine the extent and locations to which particles may be accelerated this type of model. The paper is organised as follows: in Sect.~\ref{sec:model} we discuss the model itself, introducing both the MHD simulation domain (described in Sect.~\ref{subsec:ARsim}) and the equations which govern test particle motion (in Sect.~\ref{subsec:guidingcentre}). A survey of orbit behaviour throughout the AR core is outlined and discussed in Sect.~\ref{sec:global}, before examining specific test particle examples in detail in Sect.~\ref{sec:localbehaviour}. A discussion of our findings is presented in Sect.~\ref{sec:discussion} before conclusions and future areas of study are outlined in Sect.~\ref{sec:conclusions}.

%%%%%%%%%%%%%%%%%%%%%%%%%%%%%%%%%%%%%%%%%%%
\section{Model setup}\label{sec:model}
Our model can be broadly split into two parts: a time-dependent MHD simulation (modelling the solar corona above an AR) on time-scales of hours into which we insert particles (using a single evolved snapshot), and the test particle motion itself on time scales of seconds. A brief overview of both parts are described in the following sections:

%%%%%%%%%%%%%%%%%%%%%%
\subsection{MHD active region model}\label{subsec:ARsim}

The AR MHD simulation of \citet{paper:Bourdin+al:2013_overview} that we use is observationally driven; 
\new{a potential-field extrapolation from an observed line-of-sight (LOS) magnetogram on the base determines the initial magnetic field state within the numerical domain. During the simulation the magnetic field is self-consistently computed within the domain, while the lower boundary is evolved in line with the observed time-series of LOS magnetograms.}\new{ These} magnetogram\new{s were} obtained from the {{Hinode}} solar observatory \citep{paper:Kosugietal2007,paper:Culhaneetal2007,paper:Golubetal2007}. The region in question is located close to the centre of the solar disk on the 14th November 2007; while not assigned an NOAA  AR number, the observed region displayed many typical AR characteristics (for example, a system of closed loop structures, visible in both EUV and X-ray wavelengths, linking a pair of strong magnetic field patches of opposite polarity; see Fig.~\ref{fig:Eparconfig}). 

While data from the {{Reuven Ramaty High Energy Solar Spectroscopic Imager}} \citep[RHESSI,][]{paper:Lineetal2002} was unavailable for this date, the Geostationary Operational Environmental Satellite (GOES) X-ray data suggests that solar activity at this time was particularly low. The most significant GOES event on the 14th November (and indeed for the preceding fortnight) was an A1-class event, while the emergence of a second AR (NOAA 10974) coincided with a B1-class event on the 17th November. From the available data, we characterise the chosen region as a relatively isolated, stable, and slowly evolving AR which (crucially) does not produce a flare within one day before and two days after the observation. Hence, it was an ideal choice for the original aim of the simulation, namely to study coronal heating.

The computational domain is periodic in the horizontal directions. To sufficiently isolate the main AR polarities from their periodic images, the AR observation is surrounded by a region of quiet Sun. The full horizontal extent of 235$\times$235\unit{Mm^2} is covered by 1024$^2$ grid points; approximately a quarter of this area covers the AR core (see Fig.~\ref{fig:Eparconfig}, which shows this core region alone). Thus the grid cells in the $x$$y$ directions are square with lengths of 230 km on either side.
 
The initial model atmosphere contains vertical stratification and is initialised smoothly to avoid spurious oscillations \citep[see][for more details]{paper:Bourdin+al:2014_switch-on}. Thus the domain has a stretched grid in the vertical direction that reaches up to 156\unit{Mm} above the photosphere. The vertical extent of each cell covers between 100\unit{km}, near the photosphere, up to a maximum of 800\unit{km} in the corona. For the top boundary, a potential-field extrapolation allows closed and "open" field lines to relax into a force-free state.

The compressible resistive MHD simulation was performed using the Pencil Code \citep{paper:BrandenburgDobler2002} and incorporates gravity, viscosity, and an isotropic magnetic diffusivity
$\eta (=1/\mu\sigma) = 10^{10}\unit{m^2/s}$ \citep[for electrical conductivity $\sigma$ and permeability of free space $\mu$; more details may be found in][]{paper:Bourdin+al:2013_overview}. 
\new{This value achieves diffusion of the generated current density structures at the grid scale, and is equivalent to setting the magnetic Reynolds number around unity for length scales on the order of the simulation grid scale (which is necessary to achieve numerical stability).} \new{
More details of the implementation and selection of finite resistive effects for this (and related) model(s) may be found in e.g. \citet[][]{paper:GalsgaardNordlund1996,paper:BingertPeter2011,paper:Peter2015}. We address implications of the chosen parameter regime for the recovered particle behaviour in Sect.~\ref{sec:discussion}. An image highlighting the distribution of the parallel electric field in a single snapshot of the simulation domain can be seen in Fig.~\ref{fig:Eparconfig}. Parallel electric fields are a necessary and sufficient condition for magnetic reconnection, which dissipates electric currents via Ohmic heating.}

 The initial magnetic field is evolved according to the advection of photospheric field-lines \new{using a time-series of line-of-sight magnetograms and} caused by a driver on the base which simulates horizontal granular motions in the photosphere. In this region the plasma beta is greater than one, but falls off \new{in height} as the density falls, such that the plasma beta is less than one in the corona. During the simulation run the magnetic field evolves self-consistently according to compressible resistive MHD, and subject to an energy equation that incorporates both heat conduction and radiative losses.

In particular, the energy input in the model results from the Poynting flux across the base, which provides approximately $10^7\unit{erg s^{-1} cm^{-2}}$ (vertical signed flux average) that passes through the transition-region layer and reaches the corona. This balances the expected coronal energy requirement of an AR core region by (partial) conversion of magnetic energy into heat, which forms EUV-emissive coronal loops \citep{paper:Bourdin+al:2014_coronal-loops}. A Spitzer-type heat conduction term is included in the energy equation to model energy losses in the corona due to fast electrons \citep{book:Spitzer1962}. In addition, X-ray and EUV emission is implemented through a radiative loss function which is parametrised by the local plasma density and temperature \citep{paper:Cooketal1989}.
\begin{figure*}[t]
 \centering
  \subfloat[Initial particle grid and high $E_{||}$ regions]{\label{subfig:ie}\resizebox{0.49\textwidth}{!}
  {\includegraphics{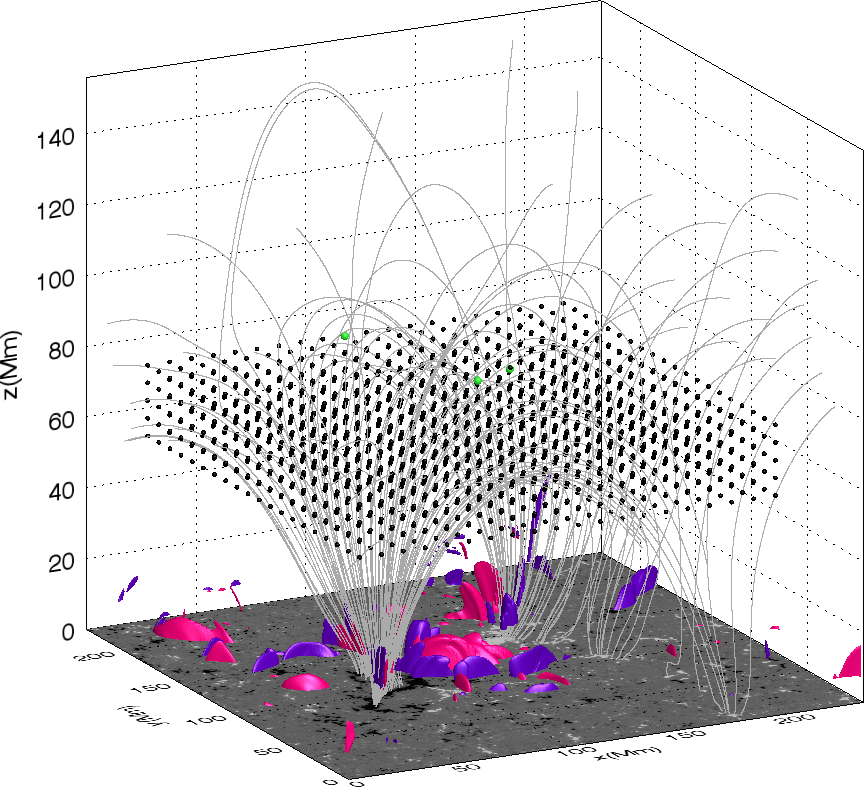}}}
  \subfloat[Moderate and high $E_{||}$ regions]{\label{subfig:ip}\resizebox{0.49\textwidth}{!}
  {\includegraphics{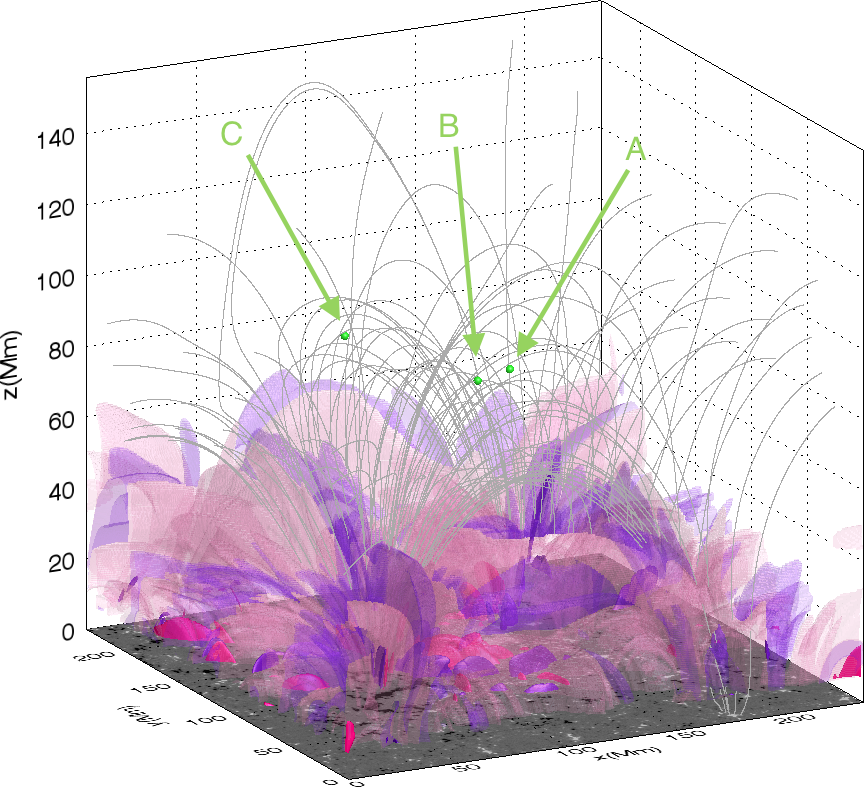}}}
 \caption{Illustration of initial particle grid and electromagnetic environment in our study. \protect\subref{subfig:ie} shows the grid of initial particle positions (shown as black orbs) located in the coronal region of MHD simulations of a stable AR \citep{paper:Bourdin+al:2013_overview}. Both \protect\subref{subfig:ie} \& \protect\subref{subfig:ip} show interpolated magnetic field lines (grey lines) and observed magnetogram at the base of the simulations, and isosurfaces of parallel electric field $E_{||}$; \protect\subref{subfig:ie} shows opaque \new{pink and purple} surfaces of \new{positive and negative $E_{||}$ respectively}, \new{encompassing regions of current} at $25\%$ of the peak value of $E_{||}$ in the corona, while \protect\subref{subfig:ip} also includes transparent surfaces of \new{$E_{||}$ at $1\%$} of the coronal peak value ($\sim3.5\unit{Vm^{-1}}$). For reference, three \new{green} orbs labelled "A", "B" and "C" indicate the initial positions of orbits studied in detail in Section.~\ref{sec:localbehaviour}.}
 \label{fig:inisurvey}
\end{figure*}
 
The lack of flaring activity and the reported agreement with observations makes this simulation an ideal candidate to study \new{particle behaviour in a realistic solar configuration}.

%%%%%%%%%%%%%%%%%%%%%%
\subsection{Relativistic particle dynamics}\label{subsec:guidingcentre}
Having established the global environment which we will study, all that remains is to outline the equations which will govern particle behaviour. Despite the lack of flaring activity, we anticipate particle velocities may achieve values of a significant fraction of the speed of light ($c$). We therefore make use of the full relativistic set of guiding-centre-motion equations, outlined in \citet{book:Northrop1963} \citep[based on the treatment of][]{paper:Vandervoort1960}, presented here in normalised form:
\begin{subequations}
 \begin{align}
  \frac{d{u_\parallel}}{dt}&=\frac{d}{dt}\left(\gamma\vpar\right)=\gamma\ue\cdot{\frac{d{\bf{b}}}{dt}}+\omscl\tscl E_\parallel-\frac{\mu_r}{\gamma}\frac{\partial{B^\star}}{\partial s}, \label{eq:Rnorm1} \\
  {\bf\dot{R}_\perp}&=\ue+\frac{\bf{b}}{B^{\star\star}}\times\left\lbrace \frac{{1}}{\omscl\tscl}\left[ \frac{\mu_r}{\gamma}\left( \grad{B^\star}+ \frac{\vsclsq}{{c^2}}\ue\frac{\partial B^\star}{\partial t}\right)\right.\right. \nonumber \\ 
   &\qquad\quad\qquad\qquad\left.\left. +u_\parallel\frac{d{\bf{b}}}{dt}+\gamma\frac{d\ue}{dt}\right]+\frac{\vsclsq}{{c^2}}\frac{u_\parallel}{\gamma}{E_\parallel}\ue \right\rbrace, \label{eq:Rnorm2} \\ 
  \frac{d\gamma}{dt}&=\frac{\vsclsq}{{c^2}}\left[\omscl\tscl\left({\bf\dot{R}_\perp}+\frac{u_\parallel}{\gamma}{\bf{b}}\right)\cdot{\bf{E}}+\frac{\mu_r}{\gamma}\frac{\partial B^\star}{\partial t}\right],   \label{eq:Rnorm3} \\
  \mu_r&=\frac{\gamma^2{\vperp^2}}{B}. \label{eq:Rnorm4}  
 \end{align}
 \label{eq:rel_norm} 
\end{subequations}
{\noindent}Here $\mu_r$ is the relativistic magnetic moment, for a particle with rest-mass $m_0$ and charge $q$, whose guiding centre is located at ${\bf{R}}$, subject to an electric field ${\bf{E}}$ and a magnetic field ${\bf{B}}$ (with magnitude $B(=|{\bf{B}}|)$ and unit vector ${\bf{b}}(={\bf{B}}/B)$). Local conditions will dictate aspects of the orbit behaviour, particularly through guiding centre drifts; the largest in magnitude is typically the ${E}\times{B}$ drift, which has a velocity $\ue(={\bf{E}}\times{\bf{b}}/B)$. The component of velocity parallel to the magnetic field is $\vpar(={\bf{b}}\cdot{\dot{\bf{R}}})$, while $\Epar(={\bf{b}}\cdot{\bf{E}})$ is the magnitude of the electric field parallel to the local magnetic field, $\dot{\bf{R}}_\perp(=\dot{\bf{R}}-\vpar{\bf{b}})$ is the component of velocity perpendicular to ${\bf{b}}$, and $s$ is a line element parallel to ${\bf{b}}$. Finally, $\gamma$ is the Lorentz factor ($\gamma^2=1/\left(1-v^2/c^2\right)=c^2/\left(c^2-v^2\right)$). Using this factor, we define a relativistic parallel velocity $\upar(=\gamma\vpar)$ for simplicity of notation. 
\begin{figure*}[t]
 \centering
  \subfloat[Final electron positions and peak energies]{\label{subfig:esurvey}\resizebox{0.425\textwidth}{!}{\includegraphics{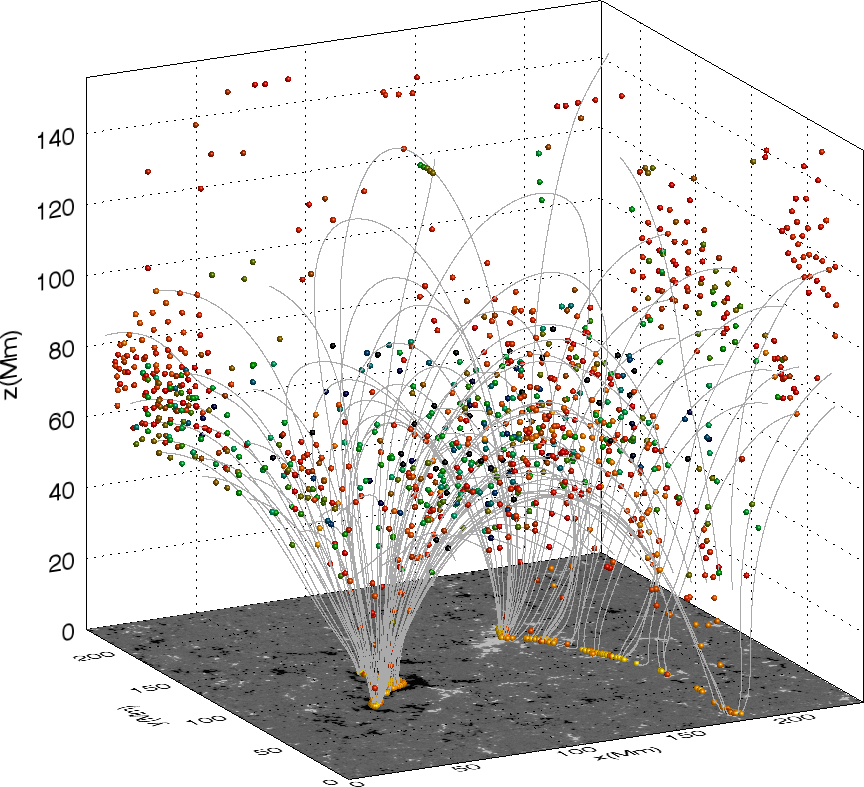}}}
  \resizebox{0.135\textwidth}{!}{\includegraphics{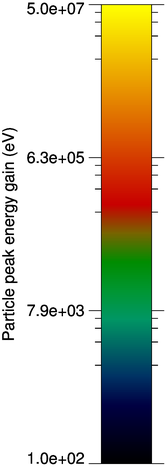}}
  \subfloat[Final proton positions and peak energies]{\label{subfig:psurvey}\resizebox{0.425\textwidth}{!}{\includegraphics{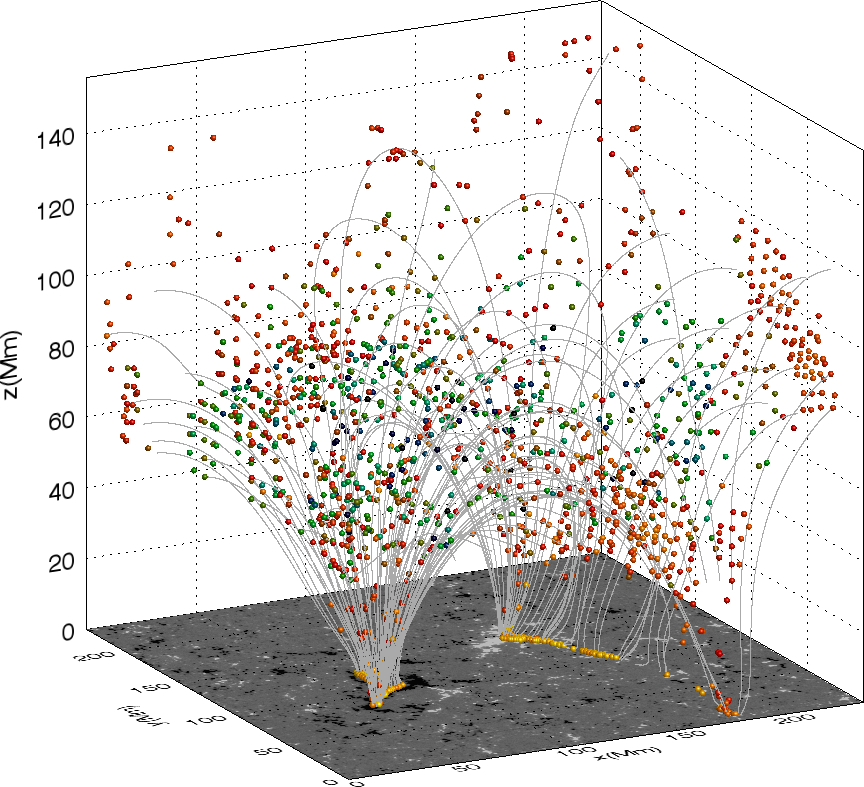}}}
 \caption{Survey of the behaviour of 2000 \protect\subref{subfig:esurvey} electron and \protect\subref{subfig:psurvey} proton orbits, for initial position grid seen in Fig.~\ref{subfig:ie}. Final particle positions are colour-coded by peak energy achieved during the orbit calculation lasting 10$\rm{s}$ (or upon leaving the MHD simulation domain) with interpolated magnetic field lines (thin grey lines) included for reference.}
 \label{fig:bigsurvey}
\end{figure*}

\new{Equations~(\ref{eq:rel_norm}) have been expressed in dimensionless form,} using a field strength $\bscl$, lengthscale $\lscl$ and timescale $\tscl$. Dimensionless quantities are related to their dimensional equivalent through:
\[
\bar{\bf{B}}=\bscl\,{\bf{B}}, \qquad \bar{x}=\lscl\,{x}, \qquad \bar{t}=\tscl\,t,
\]
where the barred quantities represent dimensional counterparts of the variables \new{seen in Eqs~(\ref{eq:rel_norm})}. This choice of quantities fixes the remaining normalising constants; for example, velocities in the model are scaled by $\vscl(={\lscl}{\tscl}^{-1})$,  energies by $\KEscl(=0.5m{\vscl}^2)$ and (assessing the dimensions of Faraday's Law) electric fields are scaled by $\escl(={\bscl\,\lscl}{\tscl}^{-1}=\bscl\,\vscl)$. This investigation is motivated by the behaviour of particles in a solar coronal environment. \new{Equations~(\ref{eq:rel_norm}) have been normalised} accordingly; unless otherwise stated, all experiments take $\bscl=0.001$\unit{T}, $\lscl=100$\unit{km} and $\tscl=10$\unit{s}.

To further simplify Eqs.~(\ref{eq:rel_norm}), only electrons or protons are considered here; this fixes the rest mass $m_0=m_e=9.1\times10^{-31}\rm{kg}$ and charge $q=e=-1.6022\times10^{-19}\unit{C}$ for electrons, or $m_0=m_p=1.67\times10^{-27}\unit{kg}$ and $q=|e|=1.6022\times10^{-19}\unit{C}$ for protons. In this way, several normalising constants are expressed in terms of a normalising electron/proton gyro-frequency, $\omscl(={q\,\bscl}{m_0}^{-1})$. The factor of $\omscl\tscl$ controls the scales at which certain guiding centre drifts become important. 

Finally, several terms in Eqs.~(\ref{eq:rel_norm}) now also depend on the ratio of perpendicular electric field ($E_\perp$) to the magnitude of the magnetic field ($B$); \new{for a given magnetic field strength $B$, $B^{\star}$ and $B^{\star\star}$ are defined as
\[
 B^\star=B\left( 1-\frac{1}{c^2}\frac{{\Eperp}^2}{B^2}\right)^{\frac{1}{2}} , \qquad B^{\star\star}=B\left(1-\frac{1}{c^2}\frac{{\Eperp}^2}{B^2}\right).
\]
(NB. $B^\star$ and $B^{\star\star}$ retain the dimensions of $B$)}.

We evolve each of Eqs.~(\ref{eq:rel_norm}) in time using a 4th order Runge-Kutta scheme with a variable timestep, subject to the numerically derived values of electric and magnetic fields found in a single snapshot of the MHD simulations of \citet{paper:Bourdin+al:2013_overview} (discussed in Section~\ref{subsec:ARsim}). A similar approach has been used by, for example \citet{paper:GordovskyyBrowning2011,paper:Gordovskyyetal2014}. Each orbit is also terminated after the normalising timescale is reached, i.e. after $10$\unit{s} of real time. The MHD simulation timescale is determined by the cadence of the observed magnetogram timeseries used to drive the photospheric boundary, which \citep[for the results reported in][]{paper:Bourdin+al:2013_overview} is $90$\unit{s}. A comparison of the time between MHD snapshots and the timescale used in the guiding centre approximation code implies that these scales are well separated. We also assume that the spatial scales of the gyro-motion and the global MHD environment are similarly separated, and will test this assumption for all orbits (see Sect.~\ref{sec:global}).

\begin{figure*}[t]
 \centering
  \resizebox{0.77\textwidth}{!}{\includegraphics{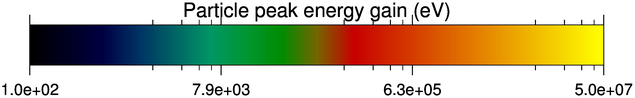}}\\
  \subfloat[${\bf{E}}=0.1{\bf{E}}_{\rm{orig}}$; electrons]{\label{subfig:esurvey10}\resizebox{0.45\textwidth}{!}{\includegraphics{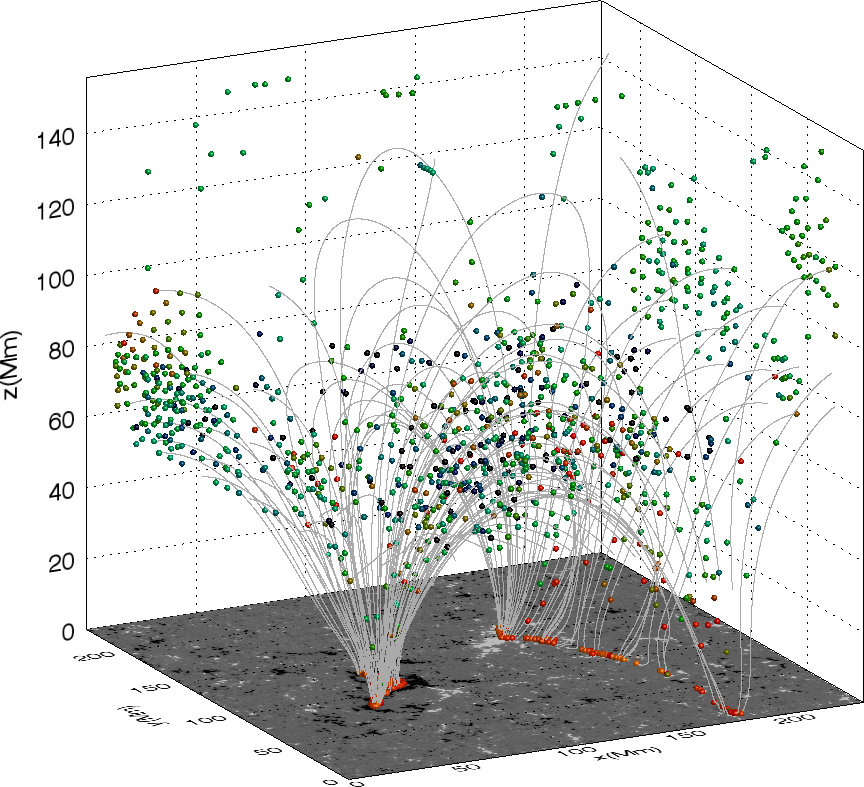}}}
  \subfloat[${\bf{E}}=0.1{\bf{E}}_{\rm{orig}}$; protons]{\label{subfig:psurvey10}\resizebox{0.45\textwidth}{!}{\includegraphics{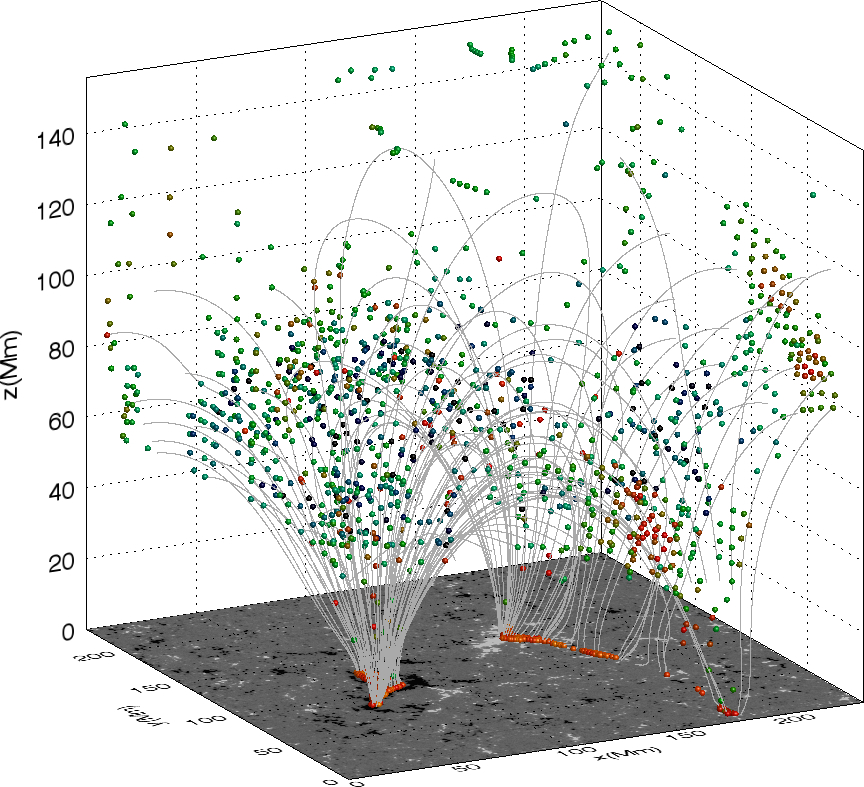}}}\\
  \subfloat[${\bf{E}}=0.01{\bf{E}}_{\rm{orig}}$; electrons]{\label{subfig:esurvey100}\resizebox{0.45\textwidth}{!}{\includegraphics{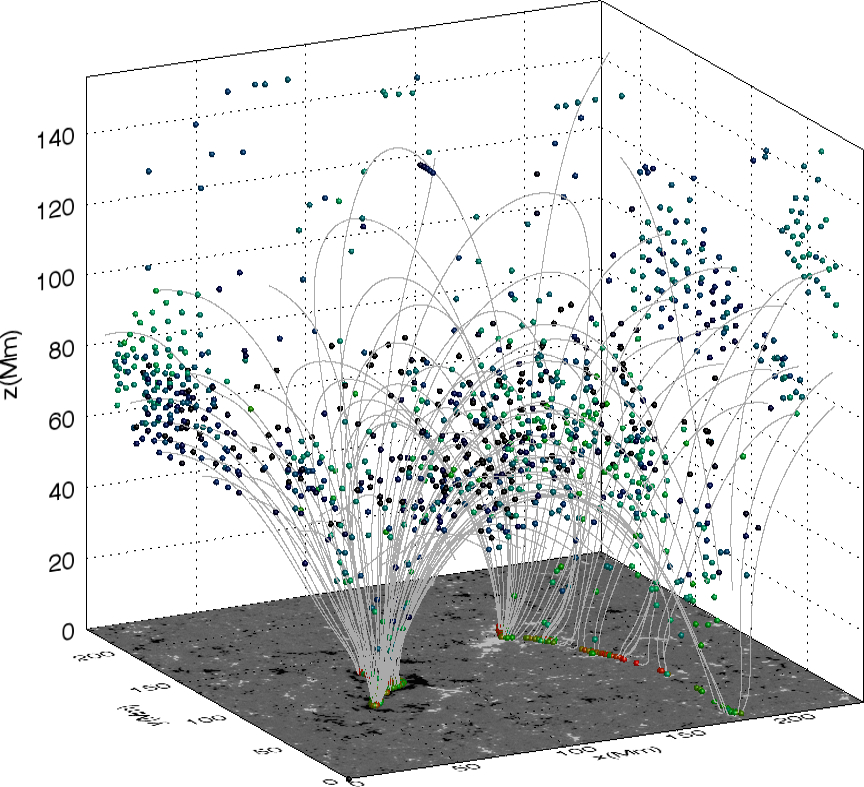}}}
  \subfloat[${\bf{E}}=0.01{\bf{E}}_{\rm{orig}}$; protons]{\label{subfig:psurvey100}\resizebox{0.45\textwidth}{!}{\includegraphics{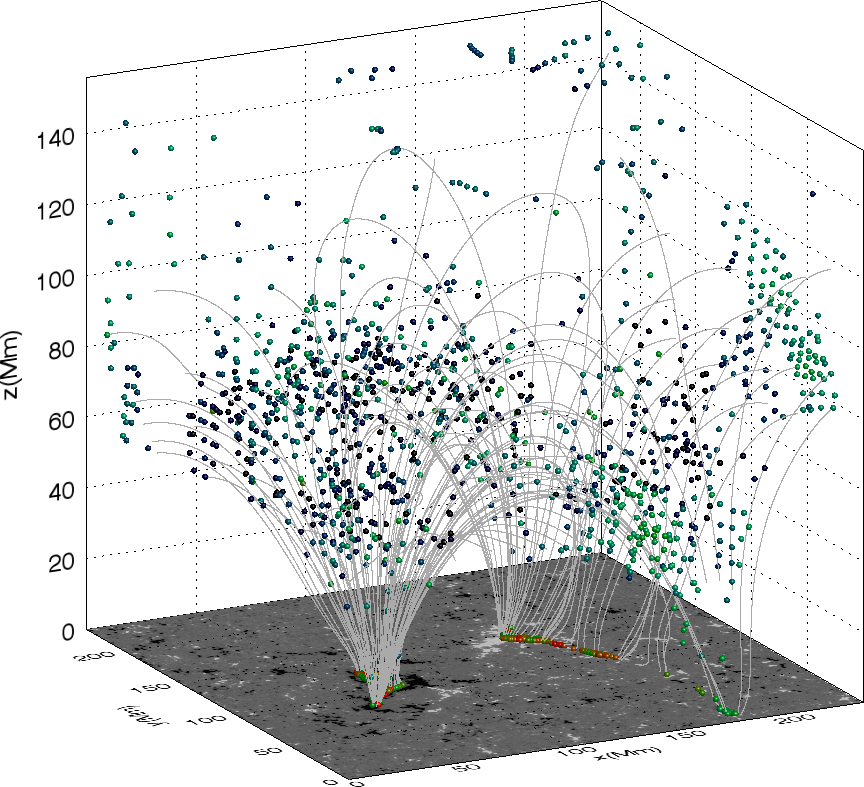}}}
 \caption{Surveys of electron and proton orbits, for initial position grid seen in Fig.~\ref{subfig:ie}, subject to electric field reductions of a factor $10$ (Figs.~\protect\subref{subfig:esurvey10}-\protect\subref{subfig:psurvey10}) and $100$ (Figs.~\protect\subref{subfig:esurvey100}-\protect\subref{subfig:psurvey100}) from the original electric field (${\bf{E}}_{\rm{orig}}$) determined by \new{the} MHD active region simulations. Final particle positions are colour-coded by peak energy \new{(using the same colour bar as in Fig.~\ref{fig:bigsurvey} to enable direct comparisons)} achieved during the orbit calculation lasting 10$\rm{s}$ for electrons or 1000$\rm{s}$ for protons (or upon leaving the MHD simulation domain) with interpolated magnetic field lines (thin grey lines) included for reference.}
 \label{fig:bigsurvey10}
\end{figure*}

%%%%%%%%%%%%%%%%%%%%%%%%%%%%%%%%%%%%%%%%%%%
\section{Characteristics of global behaviour}\label{sec:global}
%introduction
We begin by studying the particle orbit behaviour for a uniform grid of initial positions spread throughout the computational domain in the horizontal direction at several different heights. Our aim is to obtain an overview of the global orbit behaviour one might typically expect to find throughout the AR.  

%describe the initial setup of the global surveys: particle positions, energies, etc..
For this investigation, an identical number of particles for each species were studied; 2000 electron orbits and 2000 proton orbits were given an initial pitch angle of $45^{\circ}$ and an initial kinetic energy of $20$\unit{eV}. These choices are motivated by the findings of previous work \citep{paper:Threlfalletal2015}, where it was shown that (if present) the contributions from any local parallel electric fields typically outweigh the choice of initial kinetic energy and pitch angle in impacting particle behaviour. Furthermore, while the initial kinetic energy chosen here is relatively low compared to the equivalent thermal energy ($20\unit{eV}\approx0.23\unit{MK}$), this simply serves to aid illustration of particle orbit behaviour. We have also checked that a tenfold increase in initial particle energy does not significantly alter the overall findings reported here.
Our particles are distributed in a $20\times20\times5$ grid in $(x$,$y$,$z)$; the grid is evenly spaced and extends from $x,y\in[10,200]$\unit{Mm} in five $z-$planes, at $z=[60,65,70,75,80]$\unit{Mm} above the photosphere. A visual representation of this grid is shown in Fig.~\ref{subfig:ie}. This distribution of initial conditions of particle orbits was chosen to establish particle behaviour in a region encompassing the main AR features (including examples of both open and closed field structures), at large enough heights that both chromospheric effects (such as collisionality and partial ionisation) and numerical effects (primarily the inclusion of a stretched grid in $z$ close to the photospheric boundary) could be minimised. Figure~\ref{fig:bigsurvey} shows the results of both the electron and proton orbit calculations for this distribution of initial conditions.

%key findings of initial survey: high energies, trapped particles, final locations.
We will now highlight some key findings illustrated by Fig.~\ref{fig:bigsurvey}. Firstly, all orbits (both protons and electrons) are accelerated to some degree, almost always because of the local parallel electric field. The majority of orbits achieve at least keV energies; the vast majority are accelerated to energies of $50$-$500$\unit{keV}. By comparing Fig.~\ref{subfig:esurvey} with Fig.~\ref{subfig:psurvey}, it is clear that a similar number of electron and proton orbits achieve $\gtrsim$\unit{MeV} energies. All of the orbits which achieve the highest energy levels in the survey have trajectories which head towards the lower solar atmosphere. Additionally, more electron than proton orbits may be found at the lowest energy levels. It is also apparent that several orbits, both proton and electron, appear to be continually reflected between mirror points; we will label these as "trapped" orbits. 

%how many orbits are trapped?
\new{To estimate the number of trapped orbits, we count any orbit which contains two or more mirror points (locations where the parallel velocity reverses sign) as being "trapped". We estimate that, out of the $2000$ orbits initialised in the manner described earlier, $1061$ electron and $228$ proton orbits are mirrored twice or more. To present a fair comparison (with protons taking longer to reach two or more mirror points), we extended the simulation time of proton orbits \new{from 10s} to $1000\unit{s}$. While this is much longer than the $90\unit{s}$ cadence of the MHD simulations, our objective here is simply to assess under what conditions similar number of protons and electrons may be trapped. With an extended time-range, the number of trapped proton orbits rises to $840$. Thus, this trapping occurs for a significant fraction of the orbits considered in these surveys, but the true number of particles per species trapped by this mechanism is related to the time over which the MHD fields evolve, how these fields evolve, and the species mass and charge. Individual particle orbit examples, including trapped orbits, are studied in greater detail in Sec~\ref{sec:localbehaviour}.}

%put actual numbers of the kinetic energies, and check guiding centre approx is valid.
The maximum kinetic energy of any orbit was achieved by an electron, whose kinetic energy reached $41.5$\unit{MeV} (the highest energy reached by a proton was $28.7$\unit{MeV}). The minimum kinetic energy was achieved by an electron, which gained $0.09$\unit{eV} (the lowest energy gained by a proton was $0.5$\unit{eV}). Regarding the question of scale separation, the maximum electron gyro-radius recorded was $0.41$\unit{m}, while the maximum proton gyro-radius achieved during any orbit in the survey was $16.2$\unit{m}. The $x/y$ grid-point resolution of the MHD simulation is always $230$\unit{km}; the resolution in the $z-$direction varies between $100$ and $800$\unit{km} (caused by a non-equidistant grid). By comparison with the largest gyro-radii recorded, we can state that, at least in the case of the initial survey, the length scales of the MHD simulation and the particle orbits are sufficiently separate that our use of the guiding centre approximation is clearly justified.

%should I put the first bit in the discussion section?
\new{The amount of particle acceleration recovered in these orbit calculations is extreme, both in the number of particles which achieve modest/high energy gains, and the sizes of the gains themselves. The acceleration is a direct result of the electric field configuration used in the MHD active region model; the magnitudes of the electric fields are a consequence of the parameters used which were chosen such that the emission derived from the model matches the observed emission. This electric field is super-Dreicer over $40\%$ of the volume, particularly low down in the atmosphere. Super-Dreicer electric fields are known to lead to runaway particle acceleration, where collisional effects can no longer compensate for increasing particle velocities \citep[see e.g.][]{paper:Holman1985}. In these circumstances, it is perhaps unsurprising that we observe that many orbits are strongly accelerated, to highly non-thermal energies. What is surprising, though, is that not all particles which encounter strong electric fields immediately reach one of the boundaries. Many particles orbits which achieve moderate to high energies are retained at various heights in the atmosphere (resulting from particle trapping, as mentioned earlier).

% how does changing E change the surveys?
To investigate the impact of the original MHD electric field upon particle behaviour, and in light of the amount of acceleration illustrated in Fig.~\ref{fig:bigsurvey}, we repeat the orbit calculations, subject to the same initial conditions outlined earlier, but for electric fields which have been reduced by a certain factor. Figure~\ref{fig:bigsurvey10} illustrates how the peak orbit energies and final positions change when the electric field strength of the MHD simulation (which we will henceforth label as ${\bf{E}}_{\rm{orig}}$) is reduced by a factor of $10$ (for electron orbits in Fig.~\ref{subfig:esurvey10}, and proton orbits in Fig.~\ref{subfig:psurvey10}), and $100$ (electrons/protons in Figs.~\ref{subfig:esurvey100}-\ref{subfig:psurvey100}). These figures show that each decrease in electric field strength brings a corresponding reduction in the peak energy gains achieved by the orbits, while having minimal impact on the final particle positions. 

%how does changing E change the energy values and number of trapped particles?
The maximum energy gained in each experiment falls linearly with the electric field strength; peak energies in the experiment with a tenth of the original electric field strength (i.e. ${\bf{E}}=0.1{\bf{E}}_{\rm{orig}}$) were found to be $4.1\unit{MeV}$/$2.87\unit{MeV}$, exactly one tenth of the peak energies recorded in the original orbit surveys for electrons/protons respectively. This pattern continues for a hundred-fold electric field strength reduction (${\bf{E}}=0.01{\bf{E}}_{\rm{orig}}$), where the peak energies fall to $0.41\unit{MeV}$/$0.287\unit{MeV}$, \new{again} exactly one hundredth of the original peak energy values. Finally, we have once again counted the number of orbits which exhibit two or more mirror points. The trapped particle populations remain relatively consistent; out of 2000 orbits initialised in these reduced field strength cases, $821$ electron orbits remaining trapped when ${\bf{E}}=0.1{\bf{E}}_{\rm{orig}}$, and $843$ when ${\bf{E}}=0.01{\bf{E}}_{\rm{orig}}$. For protons (which were followed for 1000s), we recover $759$ and $850$ orbits in each respective field strength case which remain trapped within this electro-magnetic environment.}

To investigate the reasons behind the range of energies recovered and particle behaviours seen in this single AR system, we now focus on several examples of individual particle orbit behaviour.

\section{Examples of orbit behaviour encountered}\label{sec:localbehaviour}
A fairly broad range of behaviour was recovered in our initial survey of orbit behaviour, detailed in Section~\ref{sec:global}. We will now study specific examples of particle orbits whose initial positions are highlighted in Fig.~\ref{fig:inisurvey}; these examples illustrate different types of characteristic particle behaviour observed in the simulations, which are discussed in detail below. \new{It should be noted that, while each example highlights a single specific type of generic behaviour, the other particle behaviour may be present in each example, to a greater or lesser extent. For instance, directly accelerated particles also exhibit guiding centre drift and vice versa.}

 \begin{figure*}
 \centering
 \sbox{\bigleftbox}{%
 \begin{minipage}[b]{.48\textwidth}
  \centering
  \vspace*{\fill}
  \subfloat["Trapped" electron orbit (inc. initial/final locations, interpolated field lines, and local contour of $E_{||}=0$).]
  {\label{subfig:e25}\resizebox{\textwidth}{!}{\includegraphics{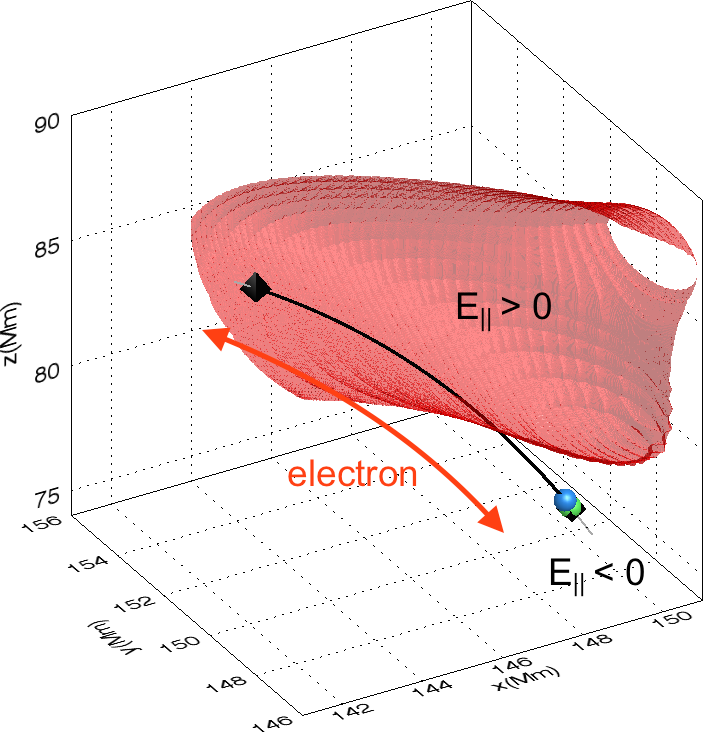}}}
 \end{minipage}%
 }\usebox{\bigleftbox}%
 \begin{minipage}[b][\ht\bigleftbox][s]{.42\textwidth}
  \centering
  \subfloat[Kinetic energy and $\Epar$]
  {\label{subfig:e25KEvEpar}\resizebox{\textwidth}{!}{\includegraphics{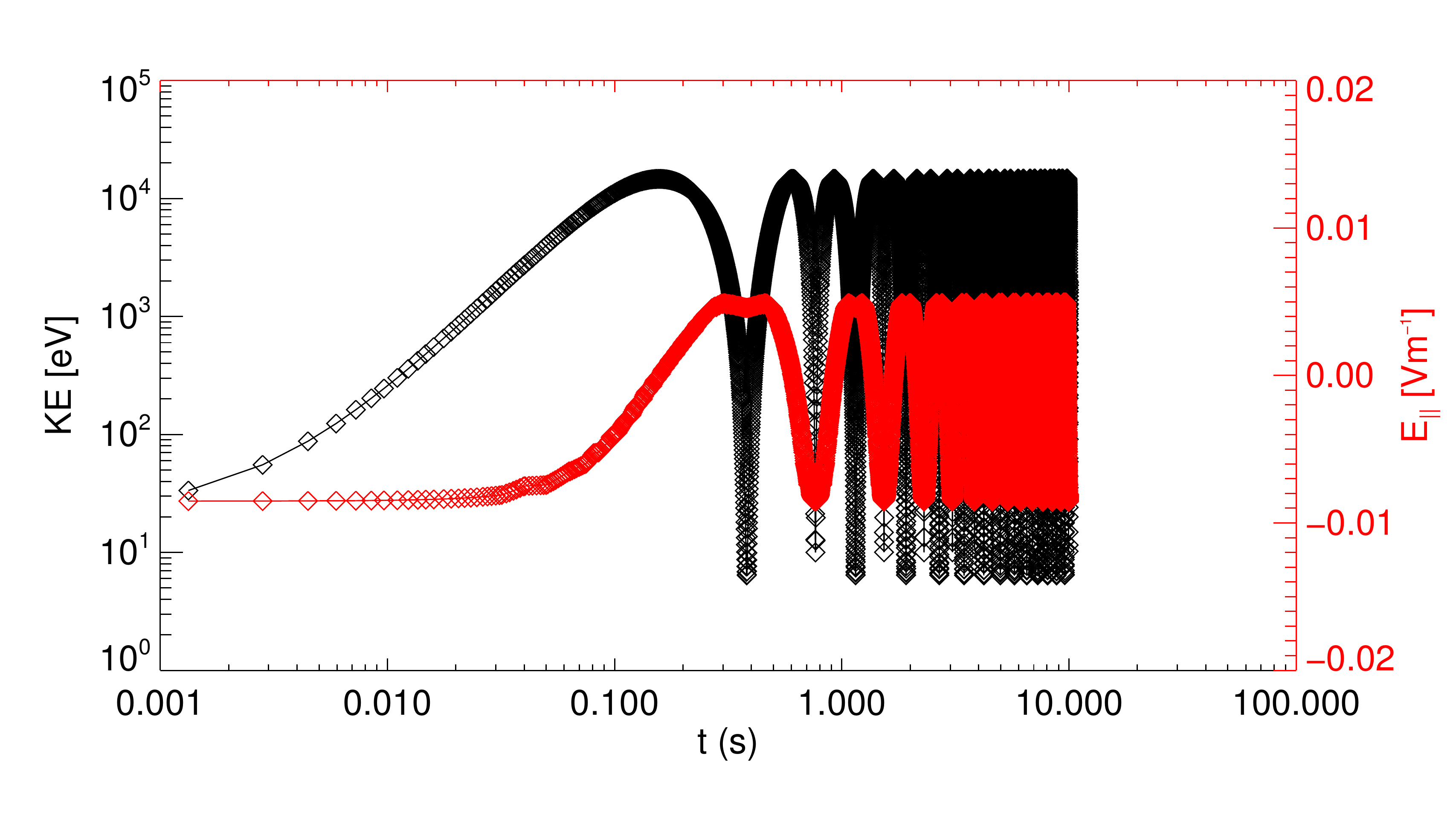}}}
  \vfill
  \subfloat[Normalised $\vpar$ and $|B|$]
  {\label{subfig:e25VvB}\resizebox{\textwidth}{!}{\includegraphics{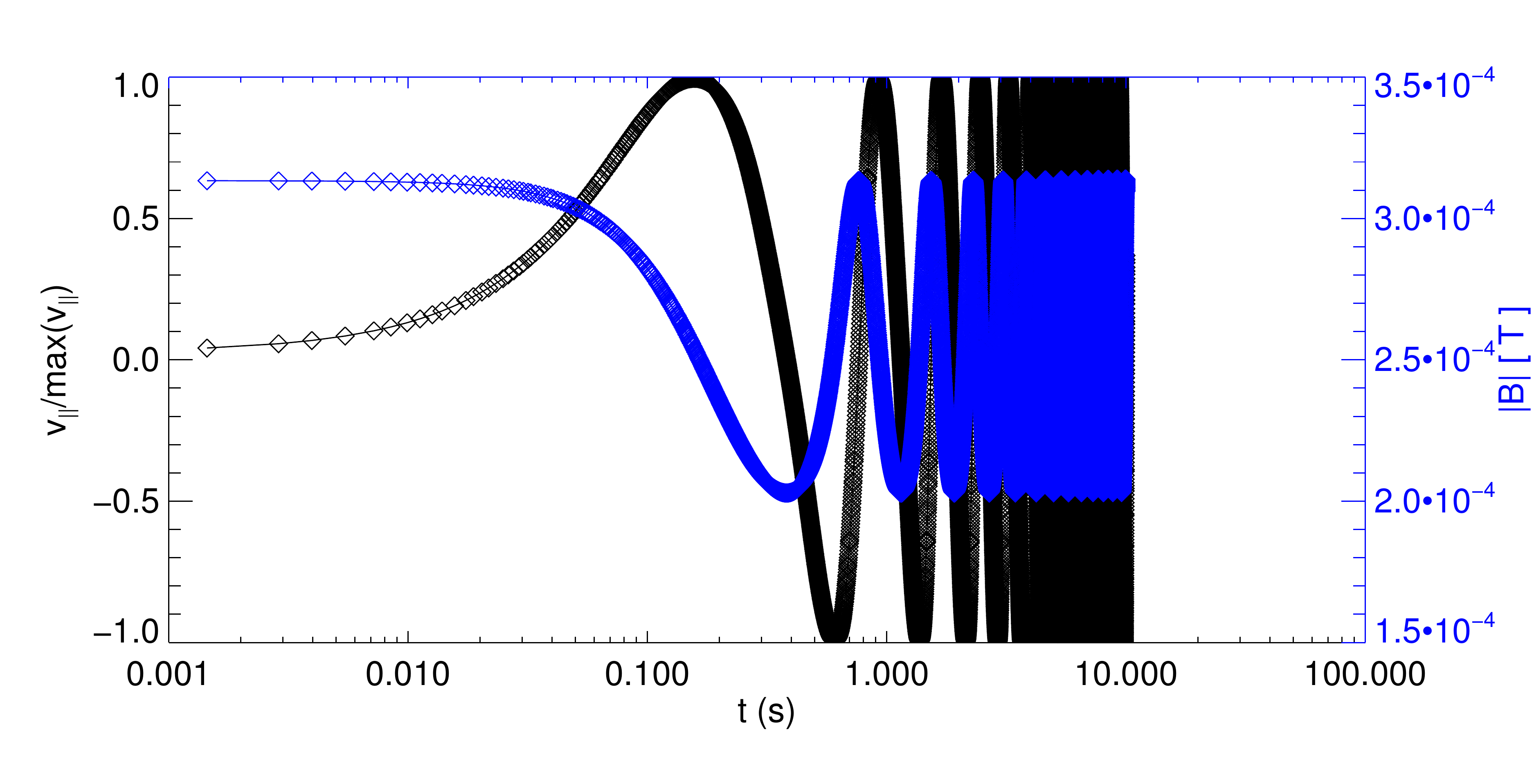}}} 
 \end{minipage}
 \caption{Position A: example of electron which becomes "trapped".  \protect\subref{subfig:e25} displays a zoomed view of the electron orbit (thick black line), which begins at the location shown by a green orb, and ends at the location shown by the blue orb, while black trapezoids indicate locations where the parallel velocity of the particle changes sign. The thin grey line indicates the locally interpolated $B$-field line based on the initial position \new{(but almost exactly matches the particle orbit path)}. The red isosurface denotes where the parallel electric field changes sign ($E_{||}=0$). Orbit properties for this electron are displayed in \protect\subref{subfig:e25KEvEpar} and \protect\subref{subfig:e25VvB}; \protect\subref{subfig:e25KEvEpar} shows the change in kinetic energy (KE) and parallel electric field component ($E_{||}$) as a function of time, while \protect\subref{subfig:e25VvB} shows the normalised parallel velocity ($v_{||}/\rm{max}(v_{||})$) and magnetic field strength ($|B|$) for this orbit.}
 \label{fig:e25}
\end{figure*}

%%%%%%%%%%%%%%%%%%%%%%
\subsection{Particle A: trapped orbits}\label{subsec:e25}
We begin with the particle labelled "A" in Fig.~\ref{subfig:ip}. This is one of many orbits which are accelerated to high (but not extreme) energies over the course of the simulation. This orbit also remains within the coronal region of the AR core throughout the orbit lifetime, as with many of the particles seen in Fig.~\ref{fig:bigsurvey}. For reference, the specific electron orbit we discuss is initially located at the centre of the grid, at position $(x,y,z)=(150,150,75)$\unit{Mm}, and remains trapped in the immediate vicinity of this position for the duration of the calculation ($10$\unit{s}), achieving a maximum kinetic energy of $14$\unit{keV} in that time. 

In Fig.~\ref{fig:e25}, we present a close-up view of this specific orbit during this time, together with a study of the local conditions (determined by the MHD simulation) encountered by the particle over the course of \new{its} %the 
orbit. From Fig.~\ref{subfig:e25}, it is clear that the particle is trapped between mirror points (locations where the local parallel particle velocity reverses sign). One common cause of particle trapping is the \emph{magnetic mirror effect} (where regions of increasing magnetic field strength cause the particle to reverse direction because of the adiabatic invariance of the magnetic moment, $\mu$). To establish the role played by the magnetic mirror effect in this particular case, we study both the local parallel velocity and encountered magnetic field strength, in Fig.~\ref{subfig:e25VvB}, which suggests that changes in sign of $v_{||}$ are not strongly associated with locations of peaks of $|B|$. To further clarify the role of the magnetic mirror effect, we calculate the loss cone angle for this position. Particles with pitch angle $\theta$ are naturally trapped by the magnetic mirror effect (in a so-called `magnetic bottle') if
\[
\sin{(\theta)}>\sqrt{\frac{B}{B_M}},
\]
\begin{figure}[t]
 \centering
  \subfloat[Parallel electric field strength]{\label{subfig:E1055}\resizebox{0.47\textwidth}{!}{\includegraphics{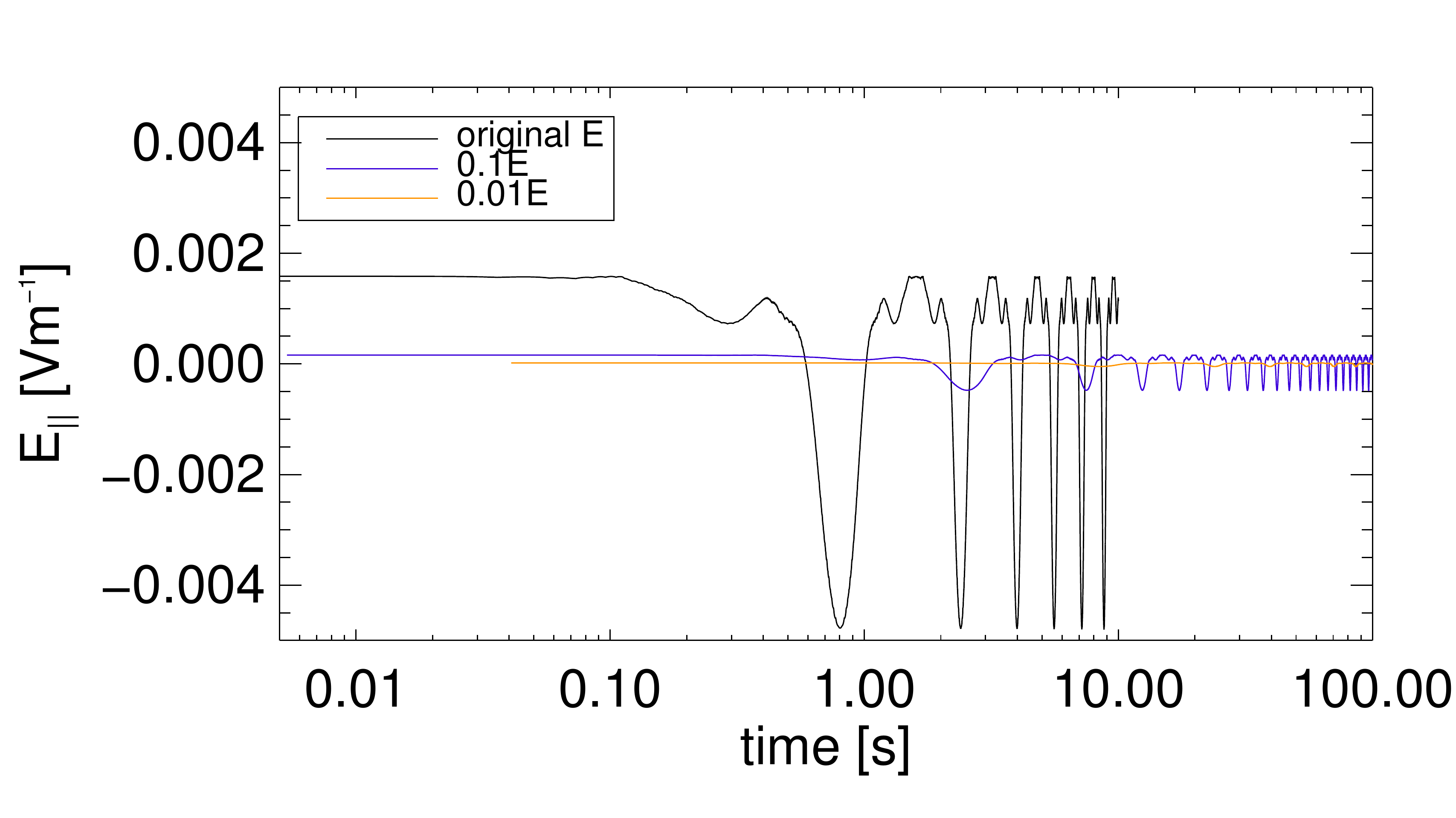}}}\\
  \subfloat[Parallel velocity]{\label{subfig:v1055}\resizebox{0.47\textwidth}{!}{\includegraphics{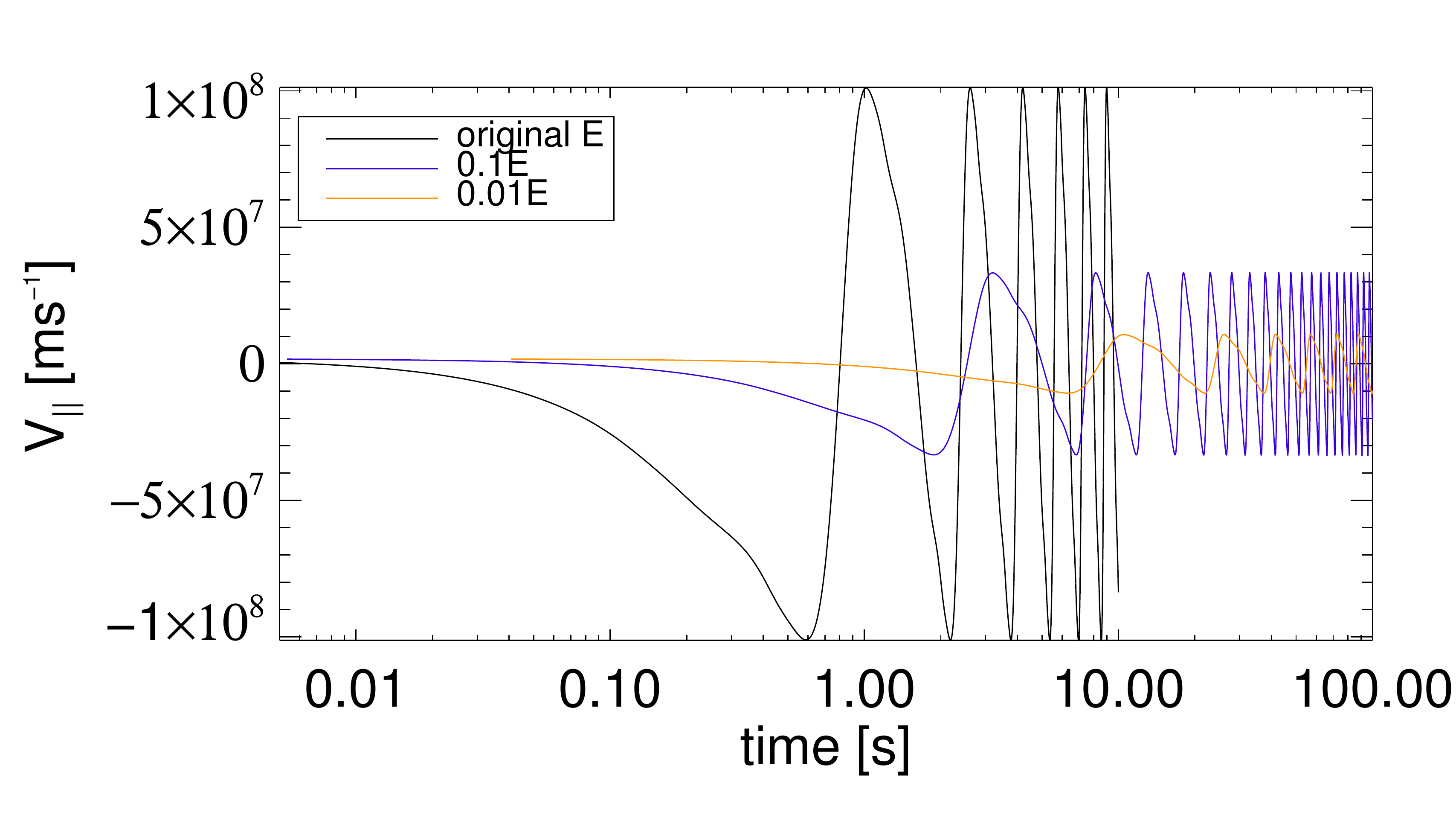}}}
 \caption{\new{Impact of decreasing electric field strength upon electric field trapping mechanism. An electron orbit (which initially is trapped between regions of strong, oppositely oriented electric field) continues to repeatedly mirror when the electric field strength is reduced by a factor of 10 (purple) and 100 (orange). \protect\subref{subfig:E1055} shows the parallel component of electric field, $E_{||}$, experienced in each field strength case (see legend) while \protect\subref{subfig:v1055} shows the corresponding changes in parallel velocity, $v_{||}$.}}
 \label{fig:changeE}
\end{figure}

where $B_M$ is the maximum magnetic field strength achieved at any location along a given field line; particles which escape this magnetic trap (i.e those with pitch angles which do not satisfy the above inequality) are said to be in the `loss cone'. The field line orbited by this electron (represented in Fig.~\ref{subfig:e25KEvEpar} by a thin grey line) is actually a closed field line, anchored in the photosphere at both ends. We estimate the peak magnetic field strength at one foot-point to be $137$\unit{G}, and $558$\unit{G} at the other; these are the locations of greatest magnetic field strength at any point along this single field line (which are of course of opposite sign in the vertical direction at the photosphere). Thus, to be in the loss cone for this particular field line, a particle must have a pitch angle $\theta<8.68^{\circ}$ in one direction and $\theta<4.30^{\circ}$ in the other. We repeated the orbit calculation for an electron with the same initial position, but with a reduced initial pitch angle of $\theta=4.0^{\circ}$; this now would ordinarily be sufficient for the particle to escape the natural magnetic trap caused by increasing magnetic field strength near the field-line foot-points. The orbit yielded by this second calculation was almost identical to that with $\theta=45^{\circ}$; the particle remains trapped regardless of initial pitch angle. We can therefore exclude magnetic mirroring as the cause of this type of electron trapping. This naturally leads to a question; if the electrons are not magnetically trapped, then how are they trapped? 
 
The answer lies in the variation of the local electric field. For this electron, the orbit path encounters a change in the sign of the local parallel electric field, $E_{||}$, as shown by the red curve in Fig.~\ref{subfig:e25KEvEpar}. An iso-surface of the contour for which $E_{||}=0$ can be seen in Fig.~\ref{subfig:e25}; the electron repeatedly crosses this contour over the course of its orbit.

Upon crossing the contour of $E_{||}=0$, the electron begins to decelerate. At the time when the local parallel velocity is approximately zero, the parallel electric field strength peaks; the orbit reverses direction (parallel to the magnetic field), re-crosses the $E_{||}=0$ contour and re-enters the region of oppositely signed $E_{||}$, leading to further (and repeated) mirroring. This pattern repeats for the entire duration of the orbit simulation. This is not an isolated example; the vast majority of electrons which do not encounter the simulation boundaries are typically "trapped" this way. This trap is particularly effective for electrons. Even relatively weak regions of oppositely signed parallel electric field strength are sufficient to repeatedly accelerate electrons this way. 

By comparison, protons are much heavier; for a proton to be trapped in this manner requires both extreme and prolonged changes in the local value of $E_{||}$ in order for the proton to be decelerated sufficiently to reverse direction. Figure~\ref{subfig:psurvey} shows that a significant number (840) of protons are also trapped, and hence are retained close to the centre of the AR core. This is unsurprising, from the size and extent of the $E_{||}$ regions near the footpoints of field-lines which thread the AR core shown in Fig.~\ref{subfig:ip}. An example of a proton which is trapped in this manner will be studied in more detail in Sect.~\ref{subsec:ep10}. 

\new{This trapping mechanism is clearly highly dependent on the (orientation of the) local electric field. However, the electric fields used in the simulation might be considered to be too strong for an active region which does not produce a flare, as they lead to highly accelerated particles. We therefore investigate how this trap is affected if the simulated electric field strength is reduced. In Fig.~\ref{fig:changeE}, we illustrate the parallel electric field strength and parallel velocity of three electron orbits for electrons that all have the same initial conditions and start from the same initial position, $(110,110,80)\unit{Mm}$, but are affected by electric fields whose strengths have been scaled to different values. Beginning with the original electric field strength, ${\bf{E}}={\bf{E}}_{\rm{orig}}$, we see from Fig.~\ref{subfig:v1055} that the orbit repeatedly mirrors, at \new{the} instants when the electric field strength peaks in Fig.~\ref{subfig:E1055}. Reducing the electric field strength by a factor of $10$ (${\bf{E}}=0.1{\bf{E}}_{\rm{orig}}$, shown in purple) shows that the orbit still repeatedly mirrors, but takes longer to reach each mirror point compared to the original field strength case. Further reducing the electric field strength
so ${\bf{E}}=0.01{\bf{E}}_{\rm{orig}}$ (shown in orange) demonstrates that this trap is 
robust; once again the orbit repeatedly mirrors until the orbit calculation is terminated, even when the orbit calculations are allowed to continue for longer timescales.}

\begin{figure*}
 \centering
 \sbox{\bigleftbox}{%
 \begin{minipage}[b]{.5\textwidth}
  \centering
  \vspace*{\fill}
  \subfloat[Drifting proton orbit (inc. initial/final locations, mirror points, interpolated field lines and contour of $E_{||}=0$).]
  {\label{subfig:p14}\resizebox{\textwidth}{!}{\includegraphics{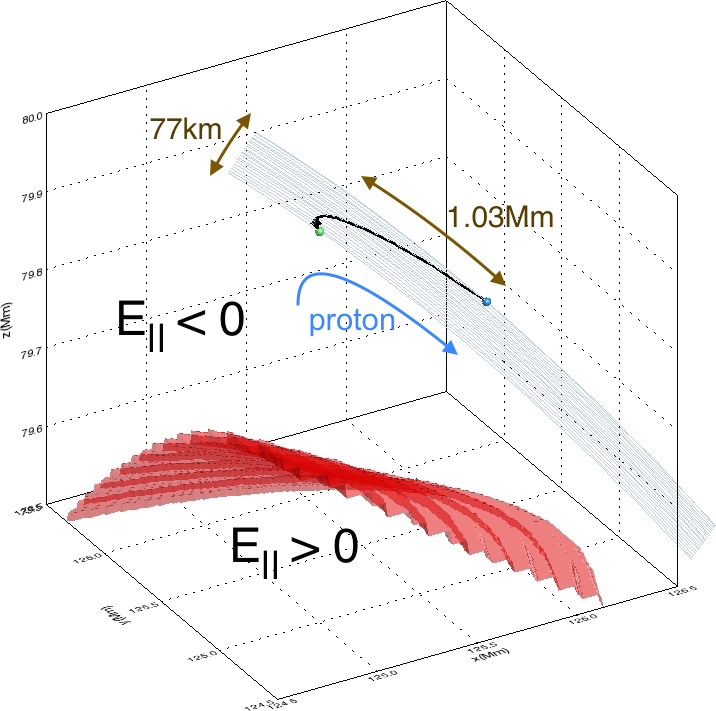}}}
 \end{minipage}%
 }\usebox{\bigleftbox}%
 \begin{minipage}[b][\ht\bigleftbox][s]{.4\textwidth}
  \centering
  \subfloat[Kinetic energy and $\Epar$]
  {\label{subfig:p14KEvEpar}\resizebox{\textwidth}{!}{\includegraphics{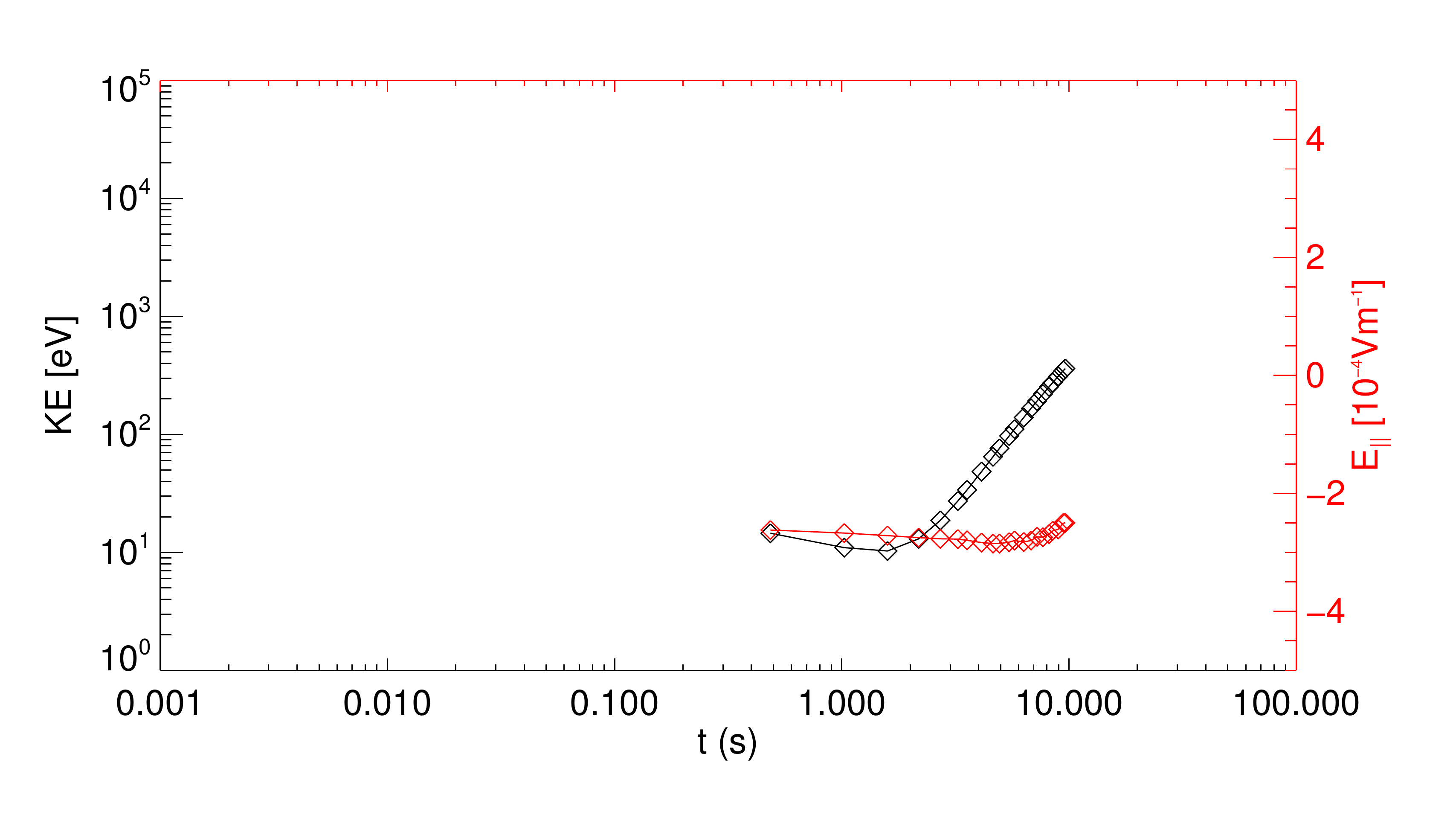}}}
  \vfill
  \subfloat[Normalised $\vpar$ and $|B|$]
  {\label{subfig:p14VvB}\resizebox{\textwidth}{!}{\includegraphics{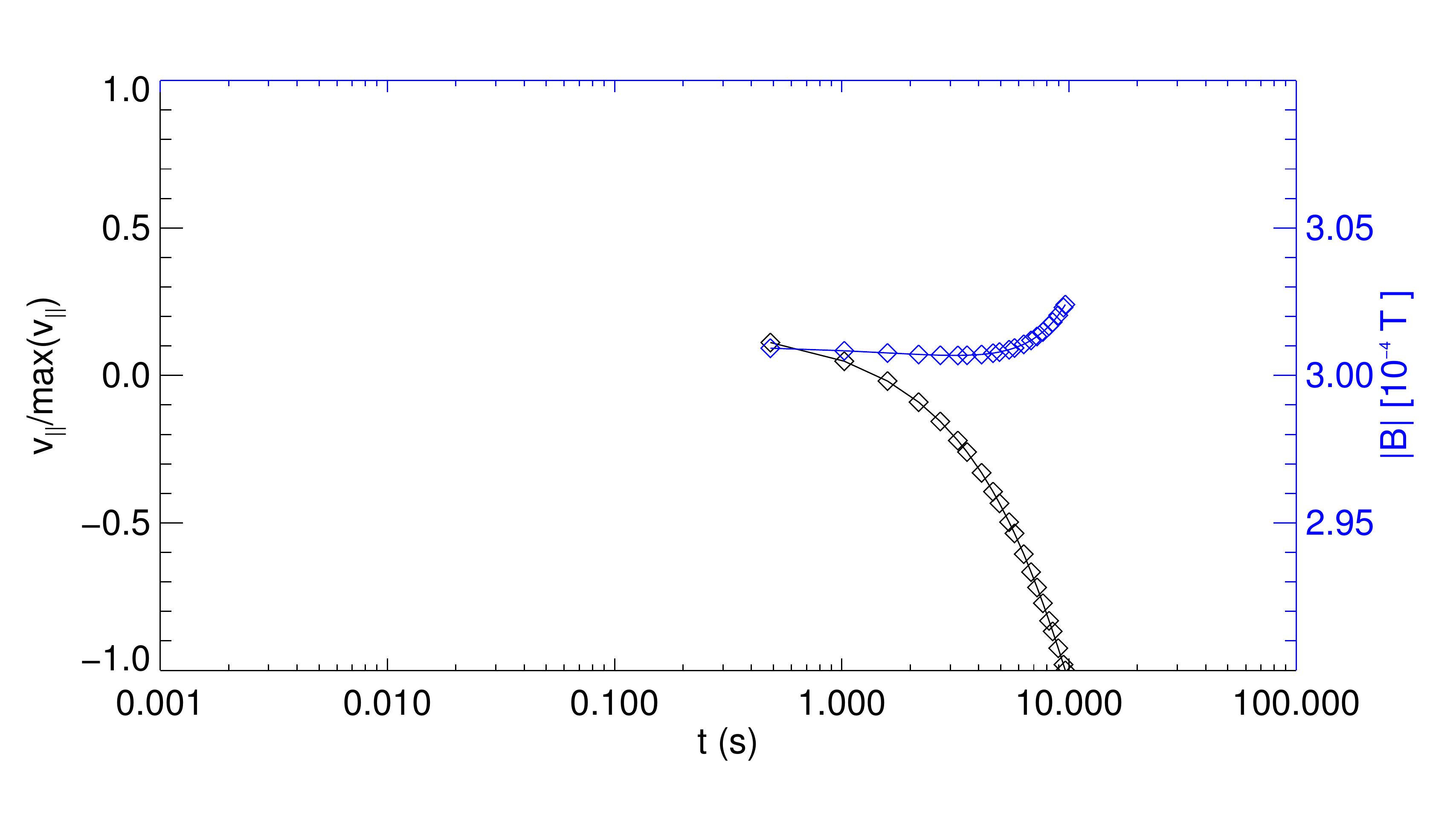}}} 
 \end{minipage}
 \caption{Position B: example of proton which exhibits a slow guiding centre drift. \protect\subref{subfig:p14} displays a zoomed view of the proton orbit (black), which begins at the location shown by a green orb, and ends at the location shown by a blue orb, with black trapezoids indicating the location of a change in sign of parallel velocity. The thin grey lines indicate locally interpolated $B$-field lines based on several orbit positions. The red isosurface denotes where the parallel electric field changes sign ($E_{||}=0$). Orbit properties for this proton are displayed in \protect\subref{subfig:p14KEvEpar} and \protect\subref{subfig:p14VvB}; \protect\subref{subfig:p14KEvEpar} shows the change in kinetic energy (KE) and parallel electric field component ($E_{||}$) as a function of time, while \protect\subref{subfig:p14VvB} shows the normalised parallel velocity ($v_{||}/\rm{max}(v_{||})$) and magnetic field strength ($|B|$) for this orbit.}
 \label{fig:p14}
\end{figure*}

%%%%%%%%%%%%%%%%%%%%%%
\subsection{Particle B: proton drifts}\label{subsec:p14}
Turning our attention to protons, we now focus on another type of behaviour seen in the initial survey. Several proton orbits are also seen \new{to %only 
gain only} a relatively small amount of energy during the $10$s calculation. One such orbit (labelled "B" in Fig.~\ref{subfig:ip}) was initialised at $(x,y,z)=(125,125,80)$\unit{Mm} and gained (at its peak) $361$\unit{eV} in kinetic energy. Once again we present a close-up view of the orbit path, together with local plasma conditions in Fig.~\ref{fig:p14}. 

It is clear from Fig.~\ref{subfig:p14} that this proton experiences a relatively constant (anti-)parallel electric field strength throughout its lifetime; the orbit path remains distant from the contour representing $E_{||}=0$ (also shown by the value of $E_{||}$ in Fig.~\ref{subfig:p14KEvEpar}). However, despite the lack of a change in sign of $E_{||}$ or a clear peak in $|B|$ (\new{see Fig.~\ref{subfig:p14VvB}) which would potentially signal the magnetic mirror effect, a} mirror point is clearly visible close to the point of initialisation (green orb), and in-between the initial and final particle positions (c.f. Fig.~\ref{subfig:e25}, where the mirror points enclose the initial and final particle positions!). The mirror point location may be understood by again considering local plasma conditions.

Having initialised the proton orbit with a positive parallel velocity (i.e. a velocity parallel to ${\bf{B}}$), the proton decelerates from its slow initialisation speed (caused by the sign of the local parallel electric field). \new{This deceleration takes place over a long timescale compared to the equivalent deceleration of an electron, resulting from the difference in species mass. During the phase where parallel orbit velocities are small,} particle drifts (primarily the $E\times B$ drift) begin to play a significant role in the proton trajectory. 

Figure~\ref{subfig:p14} includes interpolated magnetic field lines based on the changing position of the proton guiding centre over time. From this, it is clear that the proton no longer follows the same field line for all time, but instead drifts across many field lines{\footnote{The $E\times B$ drift velocity can be effectively regarded as the kinetic equivalent of the MHD fluid velocity recovered in the simulations of the (macroscopic) environmental behaviour. The fact that the $E\times B$ drift is responsible for orbits crossing many field lines in a single snapshot may initially appear to suggest \new{that a single snapshot alone may not be sufficient to accurately generate full particle orbit characteristics}.
%that the separation of MHD and kinetic scales no longer holds. 
However, the total distance drifted in the orbit simulations remains well below the spatial and temporal resolution of the MHD simulation. Thus, on an MHD level, the particle would orbit around a single field line which moves with the fluid velocity from snapshot to snapshot. However, using only a single MHD snapshot, this motion is instead seen as a drift across several (static) field lines.}}, before finally beginning to accelerate in the opposite direction to the initial direction of travel. For this particular orbit, after $10$\unit{s}, the perpendicular displacement of the proton from its original field line is approximately $77$\unit{km}; by comparison, the final parallel displacement is $1.03$\unit{Mm} (with a peak positive displacement along the field of approximately $31$\unit{km}, before being accelerated in the opposite direction over a distance of $1.06$\unit{Mm}). 
Over the course of the orbit, the proton experiences a near-constant $E\times B$ drift velocity of approximately $7.81\unit{km/s}$; integrating this value over the $10$\unit{s} orbit lifetime yields a perpendicular displacement which agrees with that observed in Fig.~\ref{subfig:p14}. Furthermore, the perpendicular displacement is well-aligned with the local ${\bf{E}}\times{\bf{B}}$ vector. While the parallel displacement is likely to further increase with time and typically will dominate over perpendicular drifts (particularly for large values of $E_{||}$ or long timescales), this case demonstrates that proton orbits may in fact drift over relatively vast distances, particularly upon encountering regions of deceleration.

\begin{figure*}
 \centering
 \sbox{\bigleftbox}{%
 \begin{minipage}[b]{.5\textwidth}
  \centering
  \vspace*{\fill}
  \subfloat[Accelerated proton \& electron orbits, initial/final locations, interpolated field lines, regions of strong $E_{||}$ and photospheric magnetogram data.]
  {\label{subfig:ep10}\resizebox{\textwidth}{!}{\includegraphics{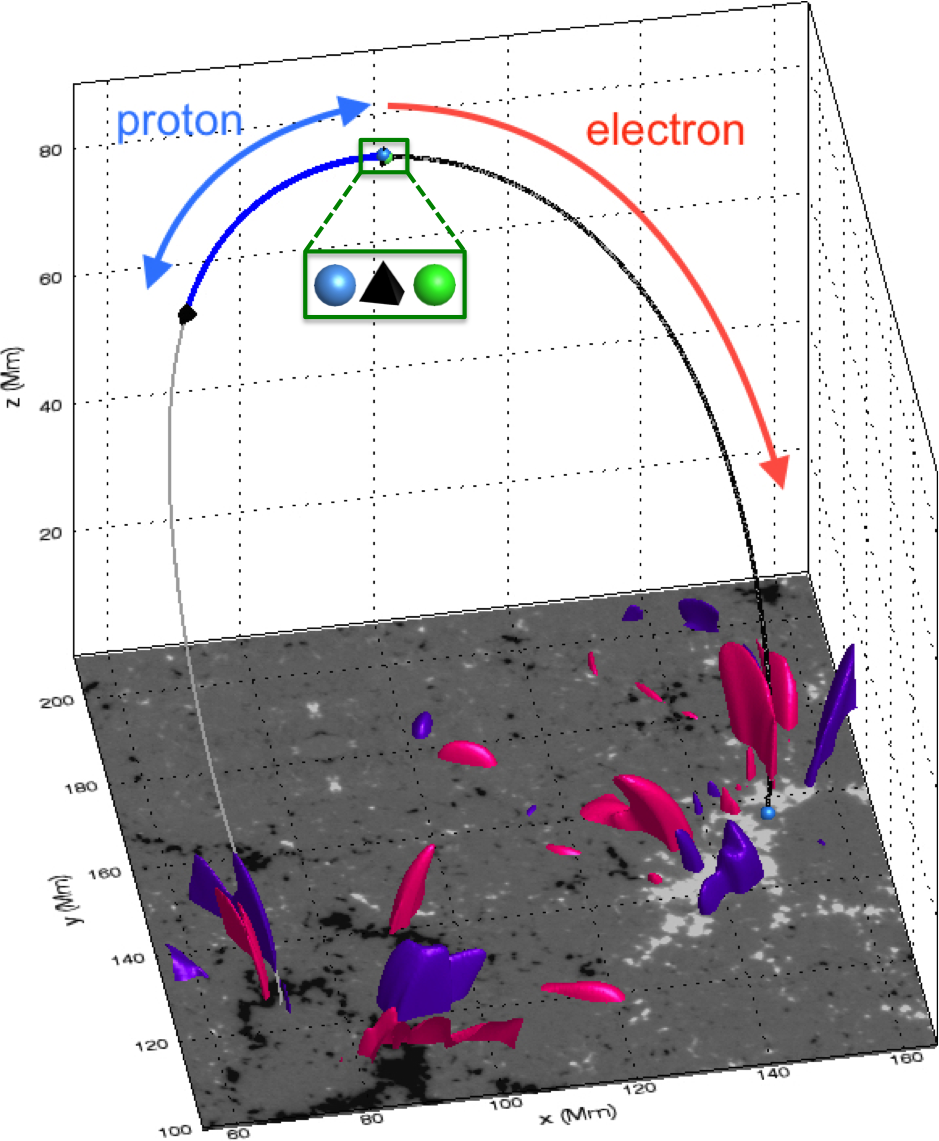}}}
 \end{minipage}%
 }\usebox{\bigleftbox}%
 \begin{minipage}[b][\ht\bigleftbox][s]{.44\textwidth}
  \centering
  \subfloat[Kinetic energy and $\Epar$]
  {\label{subfig:ep10KEvEpar}\resizebox{\textwidth}{!}{\includegraphics{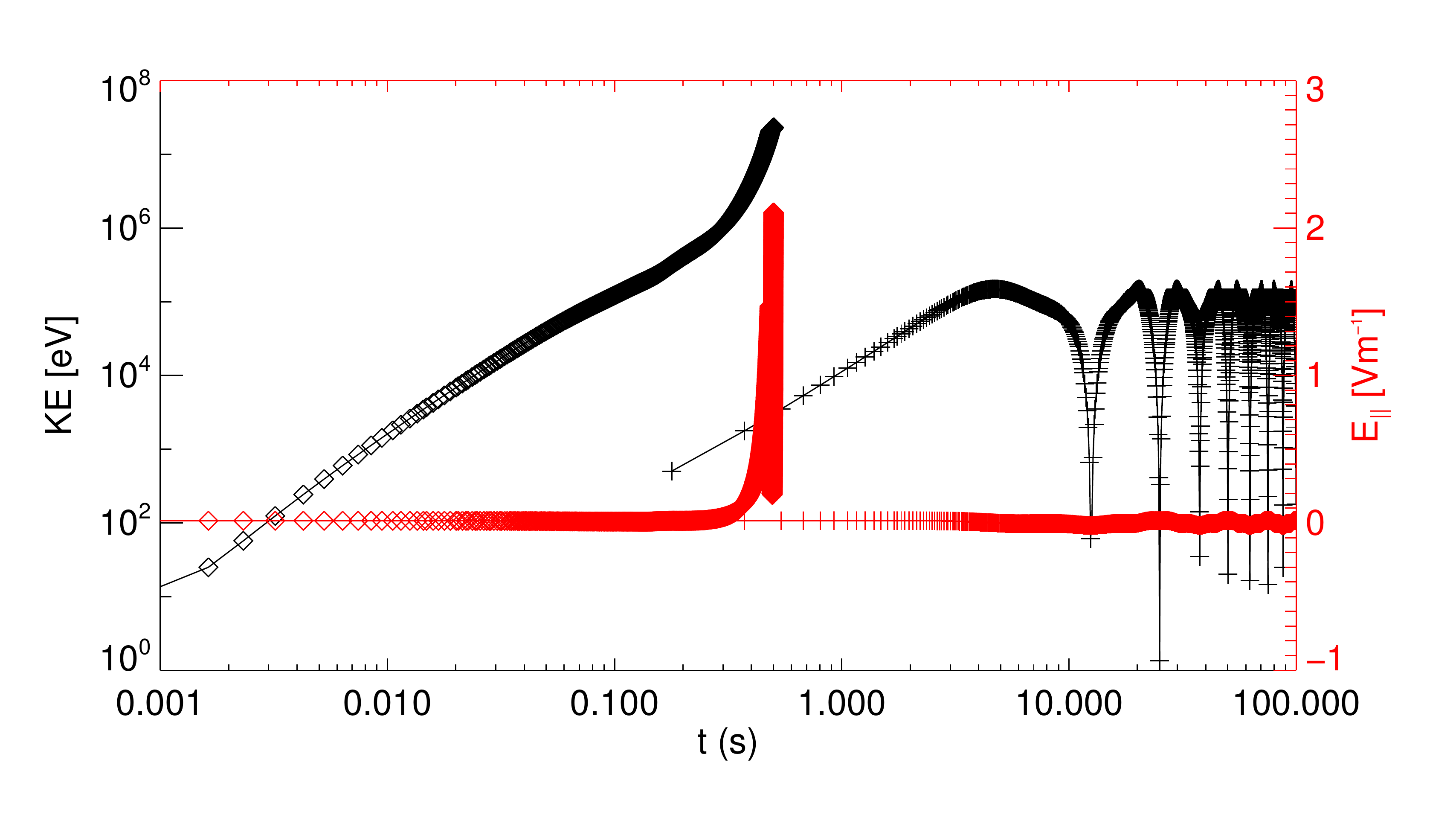}}}
  \vfill
  \subfloat[Normalised $\vpar$ and $|B|$]
  {\label{subfig:ep10VvB}\resizebox{\textwidth}{!}{\includegraphics{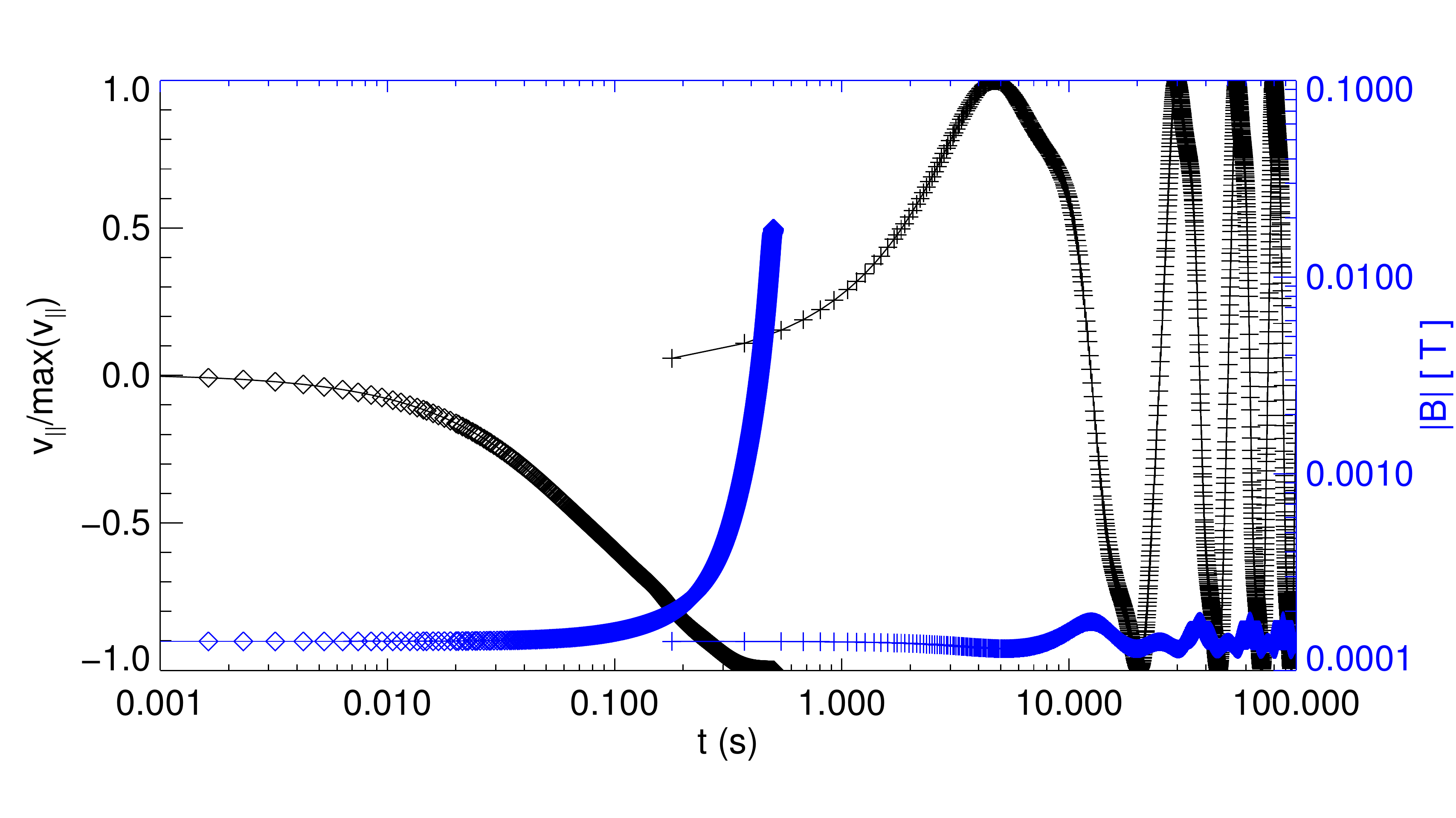}}} 
 \end{minipage}
 \caption{Position C: example of strong proton and electron acceleration. \protect\subref{subfig:ep10} displays a zoomed view of the proton (blue) and electron (black) orbits, which both begin at the location shown by a green orb, ending at a location(s) shown by a blue orb, and with changes in sign of $v_{||}$ indicated by black trapezoids \new{(noting that, on this scale, several symbols occur close together; these symbols are shown in the dark green box)}. The thin grey lines indicate locally interpolated $B$-field lines based on the initial position \new{(which, for the most part, closely match orbit paths)}, while contours of strong positive/negative $E_{||}$, and the photospheric magnetogram of vertical magnetic field strength are included for context (see Fig.~\ref{fig:inisurvey}). Orbit properties for the electron and proton are displayed in \protect\subref{subfig:ep10KEvEpar} and \protect\subref{subfig:ep10VvB} as diamonds (electron) and plus-signs (proton); \protect\subref{subfig:ep10KEvEpar} shows the change in kinetic energy (KE) and parallel electric field component ($E_{||}$) as a function of time, while \protect\subref{subfig:ep10VvB} shows the normalised parallel velocity ($v_{||}/\rm{max}(v_{||})$) and magnetic field strength ($|B|$) for each orbit.}
 \label{fig:ep10}
\end{figure*}

%%%%%%%%%%%%%%%%%%%%%%
\subsection{Particle C: Accelerated electrons/trapped protons}\label{subsec:ep10}
%Several 
Many examples of both the electron "trapping" and proton drifting can be observed in our survey. \new{However, whether or not an orbit becomes trapped is determined by the local electromagnetic configuration, the timescale over which the orbit is simulated, and the particle species. Some orbits in the initial survey are strongly accelerated in only a single direction, until they leave the simulation domain. To understand this category of behaviour, we illustrate} the orbits of both an electron and a proton which are both initialised in the same location (labelled "C" in Fig.~\ref{subfig:ip}), and which (initially) encounter (the same) strong parallel electric fields. As with previous examples, we will illustrate a close-up view of the orbit path and the local plasma parameters encountered, in Fig.~\ref{fig:ep10}. For reference, the initial position used in this example is $(x,y,z)=(100,200,80)$\unit{Mm}. To further emphasise the differences in behaviour between species, we extend the simulation time from $10$\unit{s} to $100$\unit{s} for this case. This extends the particle simulations beyond the cadence of the MHD simulations; in extending the simulation time in this manner, we simply aim to illustrate particle behaviour over a longer term.

In Fig.~\ref{subfig:ep10}, the initial position is located near the apex of a closed field line which links the foot-points of the AR studied. The proton (\new{blue} orbit) and electron (black orbit) are accelerated towards opposing foot-points (as one might expect from the charge difference between the species). Furthermore, while the electron is rapidly accelerated almost all the way to the photospheric boundary, the proton never leaves the (upper) corona, but is ultimately reflected and returns along the initial trajectory. \new{In Fig.~\ref{subfig:ep10VvB}, we see that the electron (diamonds) reaches the photosphere after approximately $0.5$\unit{s}. In this time, the proton (plus signs) has barely moved. The proton achieves maximum parallel speed after approximately $4$\unit{s} and mirrors approximately seven times in $100$\unit{s}, with the first mirroring taking place after approximately $10$\unit{s}.}  

To assess the reason for the proton becoming trapped \new{in this way}, we once again calculate the loss cone (see discussion in Sect.~\ref{subsec:e25} for more details). Along the field line in question, we estimate that initial pitch angles less than $5.81^{\circ}$ and $6.61^{\circ}$ would able particles to escape the trap caused by increasing magnetic field strength. As with the case shown in Sect.~\ref{subsec:e25}, repeating this experiment with an initial pitch angle of $4.0^{\circ}$ shows negligible differences in proton/electron behaviour; both the acceleration and trapping behaviour displayed in Fig.~\ref{fig:ep10} are caused by the electric field. In extending the simulation time from $10$\unit{s} to $100$\unit{s}, we demonstrate that the electric field continues to trap the proton in the upper atmosphere for all time, while accelerating the electron until it reaches the loop foot-point at the photospheric boundary. 

All particle orbit calculations are automatically terminated upon reaching a $5$\unit{Mm} layer above the lower simulation boundary (which results from the limited applicability of our test-particle approach in this region). At termination, the electron has travelled approximately $118$\unit{Mm} in less than $0.5$\unit{s}. By comparison, the distance between mirror points for the proton is approximately $38$\unit{Mm}. The reason for the differences between the proton and electron results are twofold; from Fig.~\ref{subfig:ep10VvB}, we can see that the proton achieves its peak velocity after approximately $4$\unit{s}, taking much longer to accelerate. Upon encountering a region of oppositely oriented $E_{||}$, the proton is unable to overcome the resulting deceleration, and mirrors. By comparison, the electron is continually accelerated as it plummets towards the photosphere. \new{The sign of the electric field strength is positive throughout the lifetime of the electron orbit (see Fig.~\ref{subfig:ep10KEvEpar}), hence the electric field trapping mechanism plays no role in the electron orbit. Furthermore, the magnetic mirror effect also plays no role in this orbit, as the strong, uni-directional electric field encountered by this electron results in a high parallel velocity ($1875\unit{km\,s^{-1}}$ or $0.006c$).}

%%%%%%%%%%%%%%%%%%%%%%%%%%%%%%%%%%%%%%%%%%%
\section{Discussion}\label{sec:discussion}
%intro 
We have conducted a survey of test-particle behaviour in an environment which aims to describe the behaviour of a solar AR. From this survey (detailed in Sect.~\ref{sec:global}), it is clear that this environment \new{(at the electric field strengths derived by the MHD model)} would cause a significant amount of particle acceleration to non-thermal energies. \new{To avoid confusion, we must first emphasise that our results are based on particle orbits and not particle fluxes. In our surveys, we have used a relatively coarse grid of initial positions to describe particle behaviour at specific locations, for a single value of initial kinetic energy and pitch angle. Such choices mean that a direct comparison with observed particle spectra from this region is not possible; we can only broadly compare common types of particle orbit behaviour.}

%clarify that we are using orbits and not fluxes - no direct comparison with spectra
\new{Even with the constraint that we cannot directly compare our findings with observed spectra, it is clear that typical orbit energy gains found in our work are much greater than would be expected from a non-flaring active region. MHD models (like the one considered here) have been shown to accurately reproduce specific observational aspects of the solar feature they have been designed to model. Our orbit calculation results show that kinetic processes also have implications for the parameter regimes chosen in such models; in our case, our orbit calculations imply that the extent of regions of strong electric field created within the MHD model would lead to significant amounts of runaway particle acceleration, which (crucially) was not observed for this specific active region. Thus particle orbit calculations could and should be used in future to inform in the design of MHD models of specific structures, with the goal of being able to closely represent the physical processes taking place within those structures, and accurately reflect observations of both heating and particle behaviour.} 

%address electric field strength issue
The \new{effects of the} strength and extent of the electric fields \new{determined in this model} are clearly illustrated in \new{the orbit results seen in} Fig.~\ref{fig:inisurvey}. We estimate the peak parallel electric field in the simulation domain to be $E_{||}^{\rm{peak}} = 295\unit{V/m}$, while in the corona (i.e. above the lowest 10 gridpoints), $E_{||}^{\rm{peak}}\approx5\unit{V/m}$. For comparison, we estimate the Dreicer electric field to approach $6\times10^{-2}\unit{V/m}$ (for $n=1\times10^{16}\unit{m^{-3}}$, $T=1$\unit{MK}). \new{Regions of super-Dreicer electric field cover extensive areas of the domain} (resulting in the orbit energies seen in Fig.~\ref{fig:inisurvey}). In \citet{paper:Threlfalletal2015}, a magnetic reconnection experiment with $E_{||}^{\rm{peak}}\approx0.1\unit{V/m}$ was sufficient to rapidly accelerate many particles to non-thermal energies, despite the apparent strength appearing too small to be considered relevant for solar flares. In fact, both the electric field strength and the \emph{extent} of the reconnection region \citep[which, for some cases in][reached $20$\unit{Mm}]{paper:Threlfalletal2015} determine the reconnection rate and (ultimately) the energy gained by the particle through direct acceleration. Thus both the strength and extent of the electric fields in the work of \citet{paper:Bourdin+al:2013_overview} are responsible for the non-thermal particle populations recovered here.

\new{We have also shown that reductions of the electric field strength \new{from} the MHD simulations (by factors of both ten and one hundred) reduce the peak energies gained by the particles by the same amount. Once again, with the energy gains directly resulting from the presence of parallel electric field, $E_{||}$, reductions in the strength of ${\bf{E}}$ reduce the voltage through which each particle may be accelerated, lowering the ultimate amount of energy they gain.}

%trapping
While many of the particles experience direct acceleration, \new{not all are accelerated out of the domain. A significant number remain trapped in the upper regions of the model atmosphere instead.} 
In Sect.~\ref{subsec:e25}, we illustrate an electron which is "trapped" between weak regions of oppositely oriented parallel electric field\new{, continuously re-accelerating to a kinetic energy of approximately $10$\unit{keV}. In Sect.~\ref{subsec:ep10}, we show an example of a proton which is similarly trapped; notably protons take much longer to be repeatedly re-accelerated, but can still achieve high kinetic energy gains ($100$\unit{keV} in the case of the proton in Sect.~\ref{subsec:ep10}) in between mirror points. The strength and extent of the two regions of oppositely oriented electric field are the primary factor that controls the peak energy which may be repeatedly gained and lost by particles of either species in an electric field trap.} 

\new{The presence of this type of trap has been noted in studies of magnetic reconnection at stochastic current sheets \citep{paper:Turkmanietal2006} and 3D magnetic null-points \citep{thesis:Stanier}\new{, but neither paper investigates the cause of the trapping}. In our work, we have studied this trap in detail, and find that the number of particles trapped by this mechanism may remain relatively unchanged despite reductions in electric field strength in the MHD model. 

For a general coronal structure, one might expect that particles are likely to encounter regions where the parallel electric \new{fields change sign} (for example between different coronal loop footpoints, or at fragmented current sheets). Our results show that this potentially may occur at many locations throughout a specific active region, and trap energised particles at heights above the photosphere, particularly in the corona. A likely way for particles to escape such a trap would be from changes in the global electro-magnetic environment; such changes would take place over much longer timescales than those considered here.}  

Whether or not this type of trapping would yield an observational signature (e.g. in radio wavelengths) is at present unknown, and may be worthy of further investigation. While trapped, these particles would also likely lose energy via collisional effects, particularly lower down in the atmosphere, or at low speeds (e.g. during deceleration/reflection). 

%other behaviour
\new{Other, more familiar types of particle behaviour are also present in our orbit surveys.} Particularly in cases which involve proton deceleration (in Sect.~\ref{subsec:p14}), particle drifts \new{may
play} a significant role in particle trajectories, \new{enabling a proton to} drift upto (potentially) tens/hundreds of kilometres from its original field line. Such drifts may contribute to the lack of alignment in X-ray and $\gamma$-ray features observed during a single solar flare event \citep[e.g.][]{paper:Hurfordetal2003}.

\new{Finally, it is worth noting that, of the characteristic types of behaviour discussed in Sect.~\ref{sec:localbehaviour}, \new{some types are more affected by} the choice of numerical resistivity used in the simulation \new{than others}. Reducing the resistivity would weaken the strength of the electric fields generated, which (in turn) reduces the magnitude of the $E\times B$ drift velocity, and the voltage differences through which particles are accelerated. However, we have shown in Sect.~\ref{sec:global} that the number of trapped orbits remains relatively constant even when decreasing electric field strengths by factors of up to $100$.} 

\new{Further investigation (Sect.~\ref{subsec:e25}) reveals that electric field strength reductions simply allow orbits to travel further prior to mirroring. \new{This results from} the orientation of the electric field, {which is unaffected by the scaling so that in weaker electric fields} orbits have further to travel before encountering \new{a} sufficiently strong electric field to be reflected, causing the majority \new{of particle orbits} to remain trapped. Indeed, with broader regions of no parallel electric field, more particles find themselves mirroring between smaller weaker regions of oppositely oriented parallel electric field. Only by encountering regions of strong $E_{||}$ (and thus achieving high velocities) or by-passing the electric field regions all-together (resulting from long-term changes in the MHD environment in which the orbits are performed) are such orbits capable of avoiding the trap set by these weak electric field regions.}

%%%%%%%%%%%%%%%%%%%%%%%%%%%%%%%%%%%%%%%%%%%
\section{Conclusions and future work}\label{sec:conclusions}
We have studied the relativistic guiding centre behaviour of test particles in an observationally driven large-scale 3D magnetic reconnection experiment \citep{paper:Bourdin+al:2013_overview}.

We find that the majority of particle motion is controlled by the strength, orientation and extent of the local parallel electric field; in \new{a significant number of cases}, the electric field causes strong direct acceleration of particles to non-thermal energies. \new{On the one hand, the strength of the electric field (which reaches super-Drecier levels throughout a significant fraction of the domain) means that collisions would play no role in the particle behaviour. On the other hand, }the lack of evidence for a flare-like event within this AR combined with our results imply that the resistivity and current structures created in the MHD simulation \new{generate too many orbits which are ultimately accelerated to unacceptably high kinetic energies. Reducing the electric field strength of the simulations leads to less strongly accelerated particles. However, even weak electric fields stretched over long distances have the capability to accelerate particles to very high energies.}

As part of this investigation, we have identified several behavioural characteristics of particles within this environment. Direct acceleration and drifting of particles were anticipated and \new{are} weakened by simulations containing weaker/shorter electric field regions. However, we also observe a type of trapping of particles between regions of oppositely oriented parallel electric field which has \new{previously only been seen in isolated, idealised studies of magnetic reconnection \citep{paper:Turkmanietal2006,thesis:Stanier}}. This type of trap results from a change in sign of the parallel electric fields, and \new{is likely to} %should 
be present in any test particle calculations based on compressible resistive MHD simulations of AR behaviour. Reducing the electric field strengths in the model does not appear to affect the efficiency of this trap. Whether or not this type of trap would produce an observational signature is unknown, but if it does it could allow for further insights into electric field structures within the solar atmosphere.

Following on from the present investigation, several opportunities for further work present themselves. The work of \citet{paper:Bourdin+al:2013_overview} merits repeating, particularly to see how the results presented here change for lower magnetic \new{diffusivities}.
The use of observations to drive MHD simulations is important, and worthy of repeating for a more recent active-region system (including regions that produce or do not produce a flare) which can be observed by a full suite of modern instruments, e.g. RHESSI, AIA/SDO, STEREO and ground-based telescopes. This might also allow for a comparison of synthetic hard X-ray data (generated by test-particle simulations) and a specific flare-like event, in a similar manner to that used in \citet{paper:Gordovskyyetal2014}. Furthermore, development of the test-particle model (for example, by studying the effect of collisions, or through the use of a distribution function which better reflects the initial state of particles in a coronal environment) would also be beneficial. 

\begin{acknowledgements}
The authors gratefully acknowledge the support of the U.K. Science and Technology Facilities Council [Consolidated Grant ST/K000950/1]. The research leading to these results has received funding from the European Commission's Seventh Framework Programme FP7 under the grant agreement SHOCK (project number 284515).
This work was supported by the International Max-Planck Research School (IMPRS) on Solar System Physics.
The results of this research have been achieved using the PRACE Research Infrastructure resource \emph{Curie} based in France at TGCC, as well as \emph{JuRoPA} hosted by the J{\"u}lich Supercomputing Centre in Germany.
Preparatory work has been executed at the Kiepenheuer-Institut f{\"u}r Sonnenphysik in Freiburg, as well as on the bwGRiD facility located at the Universit{\"a}t Freiburg, Germany.
We thank Suguru Kamio for his help finding active region observations.
Hinode is a Japanese mission developed, launched, and operated by ISAS/JAXA, in partnership with NAOJ, NASA, and STFC (UK). Additional operational support is provided by ESA and NSC (Norway).

\end{acknowledgements}

%BIBLIOGRAPHY
\bibliographystyle{aa}        % style file 
\bibliography{JT2015b}          % my .bib file
\end{document}